\pdfoutput=1
\documentclass[12pt]{article}
\usepackage{graphicx}
\usepackage[top=20mm,bottom=20mm,left=30mm,right=30mm]{geometry}

\begin{document}

\title{Scaling behavior of knotted random polygons and self-avoiding polygons: 
Topological swelling with enhanced exponent}

\author{Erica Uehara and Tetsuo Deguchi}

\maketitle

\begin{center}  
Department of Physics, Faculty of Core Research, 
Ochanomizu University \\
2-1-1 Ohtsuka, Bunkyo-ku, Tokyo 112-8610, Japan 
\end{center}

\abstract{
We show that the average size of  self-avoiding polygons (SAP) with a fixed knot is much larger than that of  no topological constraint if the excluded volume is small and the number of segments is large. We call it topological swelling. We argue an ``enhancement'' of the scaling exponent for random polygons with a fixed knot. We study them systematically through SAP consisting of hard cylindrical segments with various different values of the radius of segments.  
Here we mean by the average size the mean-square radius of gyration.  Furthermore, we show numerically that the equilibrium length of a composite knot is given by the sum of those of all constituent prime knots.  Here we define the equilibrium length of a knot by such a number of segments that topological entropic repulsions are balanced with the knot complexity in the average size. The additivity suggests the local knot picture.  
}

\newpage 
%***********************************************************************
% Section:
% Introduction
%***********************************************************************
\section{Introduction}

Ring polymers in solution have attracted much interest in various branches of science \cite{Kramers,Semlyen,Bates}.  Ring polymers with trivial topology are observed in nature such as circular DNA \cite{Vinograd}, while circular DNA with nontrivial knot types are derived in experiments \cite{Nature-trefoil,DNAknots}. Topological structures related to knots or pseudo-knots have been discussed in association with protein folding \cite{Taylor,Sulkowska}.  Naturally occurring proteins whose ends connected to give a circular topology has been recently discovered \cite{Craik}. 
Fractions of knotted species in synthesized DNA  have been measured in experiments of DNA  \cite{Rybenkov,Shaw-Wang} and recently in solid-state nanopores \cite{Plesa}. Furthermore, a molecular knot with eight crossings in nanoscale has been successfully synthesized  \cite{Woltering}.   Due to novel developments of experiments in chemistry, ring polymers are now effectively synthesized \cite{Tezuka2000,Grubbs,Takano05,Takano07,Grayson,Sugai12,Tezuka-book}.

We investigate statistical properties of knotted ring polymers in a theta solvent 
and those in a good solvent through simulations of such random polygons (RP) and self-avoiding polygons (SAP) under the topological constraint of being equivalent to a knot, respectively.  Entropic swelling occurs for the RP under a topological constraint: The average size of the RP with a fixed knot is much larger than that of no topological constraint if the number of segments $N$ is large  \cite{Deutsch,GrosbergPRL,Shimamura-JPhysA,Akos03,Matsuda,GrosbergPNAS}.  
For SAP it occurs if the excluded volume is small and segment number  $N$ is large   \cite{Miyuki-Rapid,SAP02}.  
We call such entropic swelling topological swelling \cite{Miyuki-Rapid,SAP02}. 
It has been argued that it is due to  entropic repulsions among segments of such RP or SAP with 
a knot, which were  first suggested by des Cloizeaux \cite{desCloizeaux-Let}. Here we denote by the average size the mean-square radius of gyration. 
It was shown in an experiment that the $g$-factors of ring polystyrenes, which are defined as the ratios of the mean-square gyration radii of ring polymers to those of corresponding linear polymers,  are  in theta solvents larger than the standard theoretical value and addressed that it is due to topological effects \cite{Takano12}.

In  topological swelling, it is fundamental to evaluate the scaling exponent for the RP 
and SAP with a fixed knot.  For knotted SAP  it is confirmed that  the scaling exponent  is given by that of self-avoiding walks (SAW) \cite{SAP02}, i.e., the exponent does not change. For  knotted RP  it is suggested in many simulations that the scaling exponent should be given by that of SAW \cite{Akos03,Matsuda,GrosbergPNAS}, i.e.,  
it should be enhanced due to topological swelling. 
However, it is not trivial to evaluate it numerically. 
It is practically hard to calculate knot invariants  
for polygons with large segment numbers $N$, and hence segment number $N$ is limited in the simulations  and not large enough to confirm the scaling behavior over three decades such as from $N=10^2$ to $10^5$.  It is therefore not clear yet whether the scaling exponent of the RP with a fixed knot is equal to that of SAW.

In the paper we study the scaling exponent of the RP with a fixed knot systematically  through the results of SAP by changing the excluded volume and those of  the knotting probability.   
We demonstrate that the finite-size effect is significant in topological swelling 
 both for RP and SAP and it is indeed not trivial to estimate the scaling exponent in  simulation.  However, by deriving good fitted curves to the plots of the average size versus segment number $N$  for various values of excluded volume we argue that the estimate of the scaling exponent of the SAP with a fixed knot approaches that of SAW even for zero thickness, i.e. for the RP,  as the upper limit in the plot range of $N$ goes to infinity. 

We introduce SAP consisting of $N$ impenetrable cylindrical segments with radius $r_{\rm ex}$ where the segments do not overlap each other except for neighboring ones. 
It gives a model of semi-flexible ring polymers such as circular DNA \cite{Rybenkov,Shaw-Wang}. 
We show that the scale of topological finite-size effect in segment number $N$ is given by the characteristic length of the knotting probability. 
In the cylindrical SAP it increases exponentially as a function of  the excluded-volume parameter $r_{\rm ex}$ and can be very large \cite{PLA2000,UD2015}.
Therefore, the plots of the average size of SAP with a fixed knot against $N$ do not completely show the asymptotic behavior even for large segment numbers such as $N=10^4$: the finite-size effect gradually decays over the whole plot range of $N$ for almost all SAPs investigated.   

We express the nontrivial $N$-dependence in the average size of the RP or SAP with a fixed knot 
through a formula different from the standard asymptotic expression with the fixed exponent: 
We show that a three-parameter formula with scaling exponent as a fitting parameter gives a good approximate curve to the plot of the mean-square gyration radius of the cylindrical SAP with a fixed knot against $N$ for any value of radius $r_{\rm ex}$.  
We call the parameter for the scaling exponent the effective scaling exponent. 
Since the $\chi^2$ values  are small,  the average size of the RP or SAP with a fixed knot is expressed  
as a function of both number $N$ and radius $r_{\rm ex}$ numerically with high accuracy.   
Here we remark that the mean-square gyration radius of the RP with a fixed knot was evaluated in several studies \cite{Deutsch,Shimamura-JPhysA,Akos03,Matsuda,GrosbergPNAS,Miyuki-Rapid,SAP02,Rawdon08}, and  for lattice SAP with a fixed knot the asymptotic behavior of the mean-square gyration radius was studied \cite{Marcone07}. However, such theoretical curves with small $\chi^2$ values have not been derived, yet, that have only three parameters to fit and quite small errors in the effective scaling exponent.

We shall argue the large-$N$ scaling behavior of the RP with a knot $K$ through responses in the effective scaling exponent $\nu_K$ of the cylindrical SAP with the knot $K$ as we change radius $r_{\rm ex}$ and  the upper limit in the plot range of $N$.  Here, we call it the maximum of $N$.
We show that for the cylindrical SAP of no topological constraint with a small radius $r_{\rm ex}$  the effective scaling exponent $\nu_{All}$ increases to the scaling exponent of SAW $\nu_{\rm SAW}$ very slowly as the maximum of $N$ increases. With an approximation we suggest that for $r_{\rm ex}=0.005$ the difference $\nu_{\rm SAW}-\nu_{All}$ is reduced to 0.01 when the maximum of $N$ is given by $10^8$.  We observe that effective exponents $\nu_K$  for knots $K$ are continuous as functions of radius $r_{\rm ex}$ even at  $r_{\rm ex}=0$, and also that the estimates $\nu_K$ are larger than $\nu_{All}$ for any value of $r_{\rm ex}$. We thus suggest that effective scaling exponent $\nu_K$ of the RP with a knot $K$ approaches $\nu_{\rm SAW}$ if the maximum of $N$ is very large such as $10^8$.

As another aspect of topological swelling we shall show the additivity of equilibrium lengths for composite knots.  We call the number $N_{\rm eq}$ the equilibrium length of a knot $K$, if the mean-square radius of gyration of such SAP of $N_{\rm eq}$ segments with a knot $K$ is equal to that of  such SAP of $N_{\rm eq}$ segments with no topological constraint.  We may interpret that when the segment number  is equal to the equilibrium length 
the average size does not change even if we remove the topological constraint on the SAP \cite{Rawdon08}. Intuitively, in the RP or SAP with a fixed knot at the equilibrium length  
entropic repulsions arising from the topological constraint are balanced with the complexity of the knot. Recall that the average size of knotted SAP decreases for small $N$ if the knot is more complex.   
We propose a conjecture that for a composite knot $K_1 \# K_2$ which consists of knots $K_1$ and $K_2$, the equilibrium length $N_{12}$ of  knot $K_1 \# K_2$ is given by the sum: $N_{\rm eq1} + N_{\rm eq 2}$,  where $N_{{\rm eq} j}$ are the equilibrium lengths of knots $K_j$ for $j=1$ and $2$. For the  cylindrical SAP we show that the conjecture holds for SAP, while for RP it holds but not as good as for SAP:  $N_{12}$ is smaller than the sum for RP. 
We shall argue that  the additivity of equilibrium lengths 
is consistent with the local knot picture that the knotted region in a knotted SAP is localized \cite{Orlandini1998,Katritch00,Marcone05,Rosa11}.  It  might be suggested that the knotted region is less localized in RP than in SAP.

The contents of the present paper consist of the following. 
In section 2 we explain numerical methods in this research. 
In section 3 we present several aspects of topological swelling. 
We first define the ratio of topological swelling for a knot $K$ by the ratio of the mean-square radius of gyration of the RP or SAP with the knot $K$ to that of no topological constraint. By plotting the ratio of topological swelling for several knots against segment number $N$  
we demonstrate that the finite-size effect is dominant in the plots. 
We show that all the data of the ratio of topological swelling are well approximated by the fitted curves derived from the three-parameter formula with an effective scaling exponent applied to various knots for many different values of cylindrical radius $r_{\rm ex}$. 
In section 4, we show that the three-parameter formula with scaling exponent as a fitting parameter gives good fitted curves to the data points of the mean-square gyration radius of SAP with a knot $K$, denoted by $\langle R^2_g \rangle_K$, against segment number $N$ for various values of radius $r_{\rm ex}$.  We obtain good fitted curves to five  prime knots and five composite knots such as  
$0_1$, $3_1$, $4_1$, $3_1\# 3_1$ and $3_1 \# 4_1$, etc. 
Here we remark that we define a prime knot as such a knot that cannot be expressed as 
a product of two nontrivial knots \cite{Murasugi}.   In section 5 we argue that the effective scaling exponent $\nu_K$ of the RP with any fixed knot $K$ approaches  the scaling exponent of SAW  (i.e., $\nu_{\rm SAW}$=0.588) if the upper limit in the plot range of segment number $N$ goes to infinity. We point out that the topological finite-size effect is very strong, so that the estimate $\nu_K$ of the RP with knot $K$ is smaller than the scaling exponent of SAW $\nu_{\rm SAW}$ if the number of segments $N$ is less than $10^4$.  In section 6 we present the additivity of equilibrium lengths  
for several composite knots consisting of prime knots $3_1$ and $4_1$.  
Finally, in section 7 we give concluding remarks.

%***********************************************************************
% Section 2
% 
%***********************************************************************
%
\section{Numerical methods}

\subsection{Algorithm for generating cylindrical SAP}
We construct an ensemble of SAP consisting of $N$ hard cylindrical segments with radius $r_{\rm ex}$ as follows \cite{UD2015}. First, we construct an initial polygon by an equilateral regular $N$-gon, where the vertices have numbers from $1$ to $N$, consecutively. Second, we choose two vertices randomly out of the $N$ vertices. Suppose that they are given by numbers $p_1$ and $p_2$.We rotate the sub-chain between the vertices $p_1$ and $p_2$ around the straight line connecting them by an angle chosen randomly from $0$ to $2\pi$. Third, we check whether the rotated subchain has any overlap with the other part of the polygon or not. If the distance between every pair of non-neighboring segments (or polygonal edges) of the polygon is larger than $2 r_{\rm ex}$, we find that the polygon does not have any overlap. If it has no overlap, we employ the rotated configuration as the cylindrical SAP in the next Monte-Carlo step. If it has an overlap, we employ the previous configuration of SAP before rotation in the next Monte-Carlo step. Then, we repeat this procedure many times such as $2N$ times.

When the correlation between the initial polygon and the current polygon becomes small, we add the current polygon to the ensemble of cylindrical SAP. Repeating the procedure $2N$ times makes the correlation small enough.  We thus suggest that if we pick up a SAP in the employed configurations every $2N$ Monte-Carlo steps, the correlations between samples are very small and can be neglected.

\subsection{On the number of SAPs generated in simulation}

In the simulation of the present paper to each value of cylindrical radius $r_{\rm ex}$ we have generated  $2\times10^5$ polygons for $N \le 4000$,  
$10^5$ polygons for  $4000 < N \le 6000$,  
$5 \times10^4$ for  $6000 < N \le 8000$, and 
$ 4 \times10^4$ for $8, 000 < N \le 10, 000$.

We remark that for $r_{\rm ex}=0$ generated polygons in the cylindrical SAP model 
are given by equilateral random polygons.

In the present simulation the number of segments $N$ is given from $100$ to $3000$ for the cylindrical SAP of zero thickness ($r_{\rm ex}= 0$),  
from $100$ to $3000$ for cylindrical SAP with  $r_{\rm ex}=0.005$ and $r_{\rm ex}=0.01$, from $100$ to $4000$ with $r_{\rm ex}=0.02$, from $100$ to $5000$ with $r_{\rm ex}=0.03$, from $100$ to $7000$  with $r_{\rm ex}=0.04$, from $100$ to $8000$ with $r_{\rm ex}=0.05$, from $100$ to $10^4$ with $r_{\rm ex}=0.06$, 0.08 
and 0.1.

In order to detect the knot type of a given SAP we  mainly evaluate the two knot invariants: The absolute value of the Alexander polynomial $|\Delta_K(t)|$ evaluated at $t=-1$ and the Vassiliev invariant of the second order $v_2(K)$ for a knot $K$ 
\cite{Deguchi-Tsurusaki-PLA}. 
If a given SAP has the same values of two knot invariants as a knot $K$, 
we assume that the topology of the polygon is given by the knot $K$. 
However, for some cases we also evaluate the Vassiliev invariant of the third order, 
In fact, there are some known pairs of knots that have the same values of the two knot invariants.
We employ the algorithm for calculating the Vassiliev invariants of the second order and the third order 
 through the Gauss codes (or the Dowker codes)  \cite{Polyak-Viro}.

%%%%%%%%%%%%%%%%%%%%%%%%%%%%%%%%%%%%%%%%%%%%%%
% section 3 
%
\section{Entropic swelling of SAP with a fixed knot }

%%%%%%%%%%%%%%%%%%%%%%%%%%%%%%%%%%%%%%%%%
\subsection{Ratio of topological swelling}  
%: $\langle R^2_{g} \rangle_{K}$/$\langle R^2_{g} \rangle_{All}$}

We denote the mean-square radius of gyration of the cylindrical SAP with a knot $K$ by $\langle R^2_{g} \rangle_{K}$ and that of no topological constraint by  $\langle R^2_{g} \rangle_{All}$. 
Here we recall that the cylindrical SAP with zero radius: $r_{\rm ex}=0$ correspond to 
equilateral RP.

%-----------------------------------------------------------------------
% Figure 1. Ratio 01knot 
\begin{figure}[htbp]
\begin{center}
 \includegraphics[width=
%12.5cm
0.8\hsize]{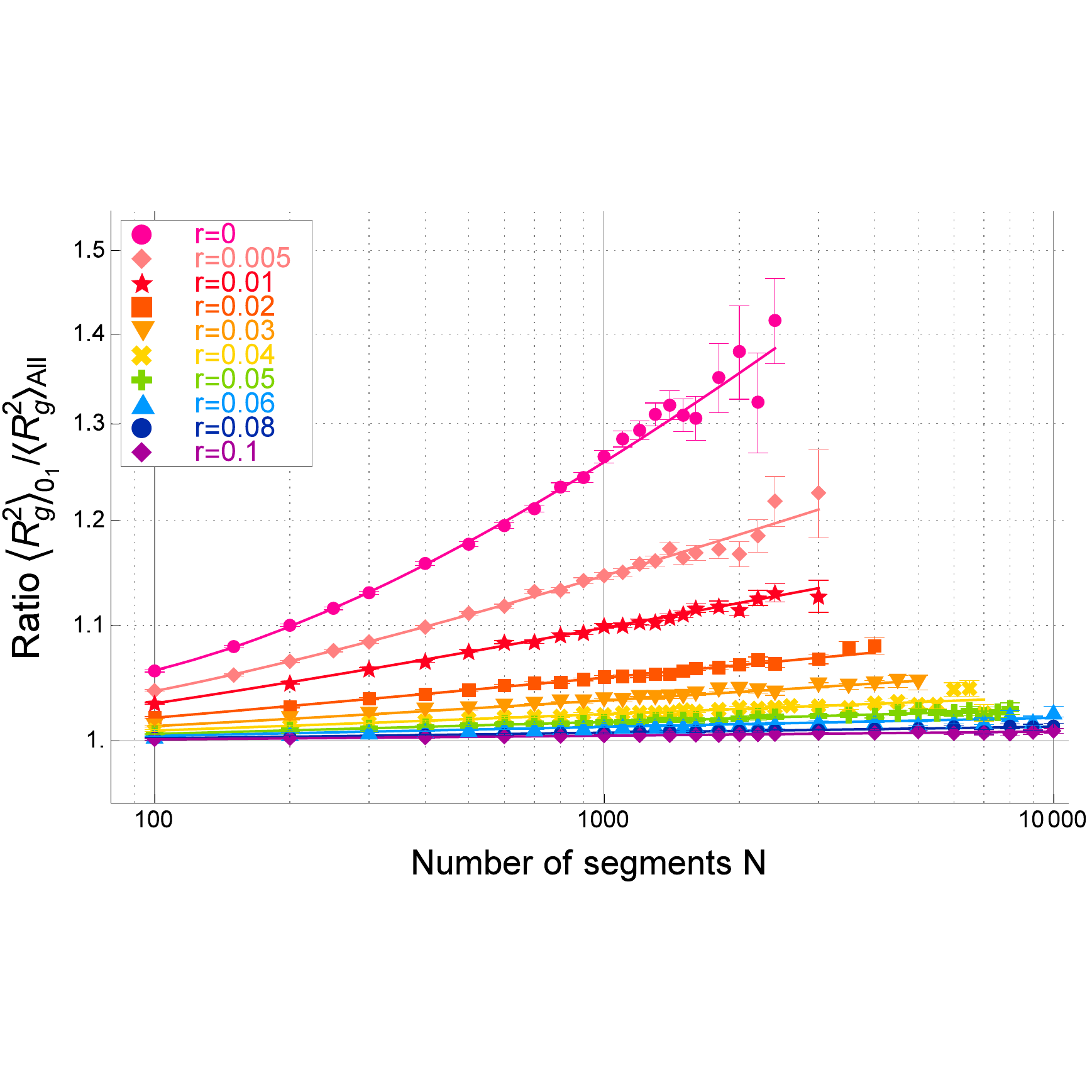}
  \caption{Ratio of topological swelling for the trivial knot ($0_1$) in the cylindrical SAP with radius $r_{\rm ex}$,   $\langle R^2_g \rangle_{0_1}/\langle R^2_{g} \rangle_{All}$ plotted against the number of segments $N$ in the double-logarithmic scale.  The plots for the values of radius $r_{\rm ex}$ given by 0, 0.005, 0.01, 0.02, 0.03, 0.04, 0.05, 0.06, 0.08 and 0.1 are depicted by circles (red), diamonds, stars,  squares,  lower triangles, saltires or Xs, crosses, upper triangles, circles (blue) and diamonds (purples), respectively.
}
  \label{fig:ratio01}
\end{center}
\end{figure}
%-----------------------------------------------------------------------

\begin{figure}[htbp] \begin{center}
 \includegraphics[width=
%12.5cm
0.8\hsize]{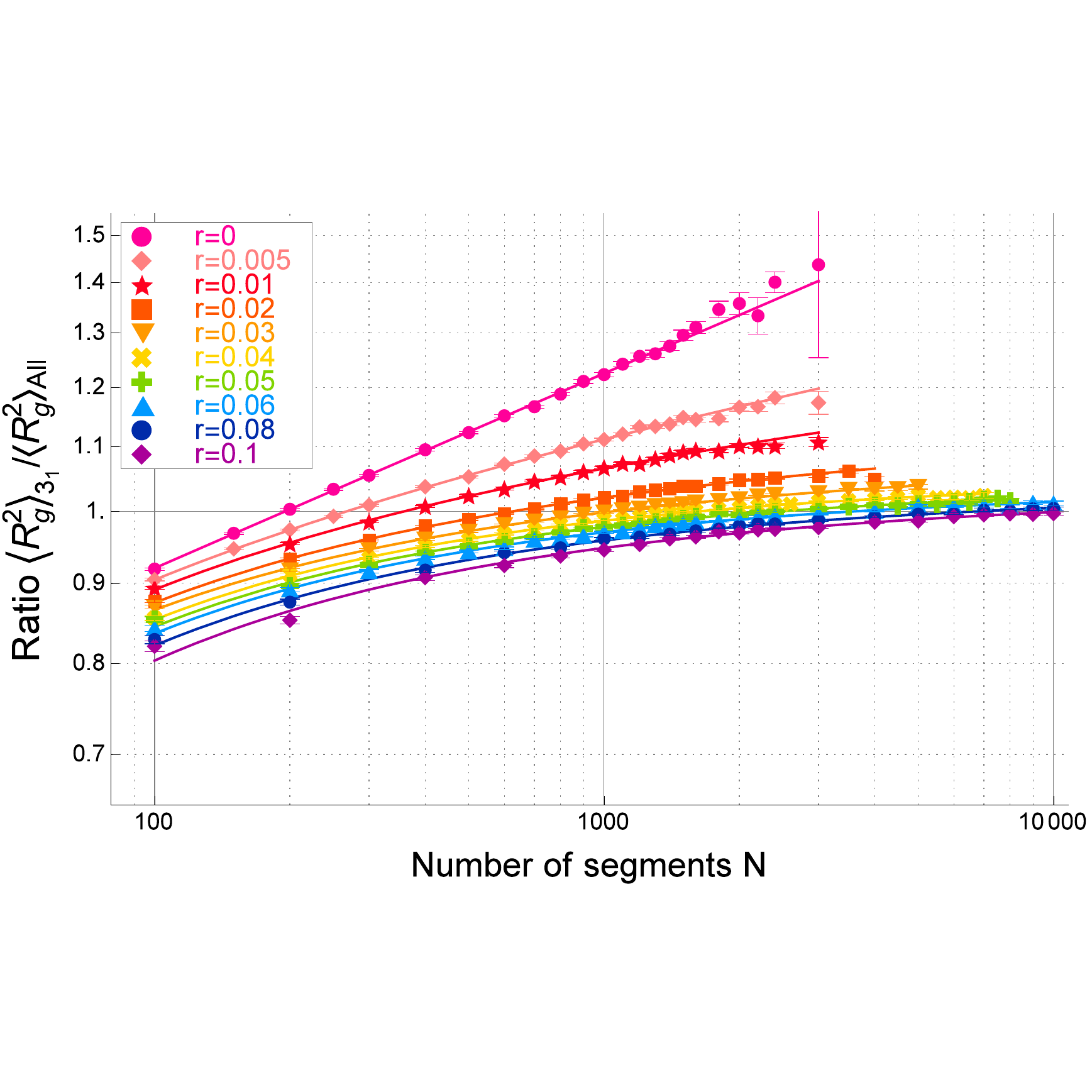}
 \caption{Ratio of topological swelling for the trefoil knot ($3_1$) 
in the cylindrical SAP with radius $r_{\rm ex}$,   
$\langle R^2_g \rangle_{3_1}/\langle R^2_{g} \rangle_{All}$ plotted against the number of segments $N$ in the double-logarithmic scale. } \label{fig:ratio31}
\end{center} \end{figure}

\begin{figure}[htbp]
\begin{center}
\includegraphics[width=
%12.5cm
0.8\hsize]{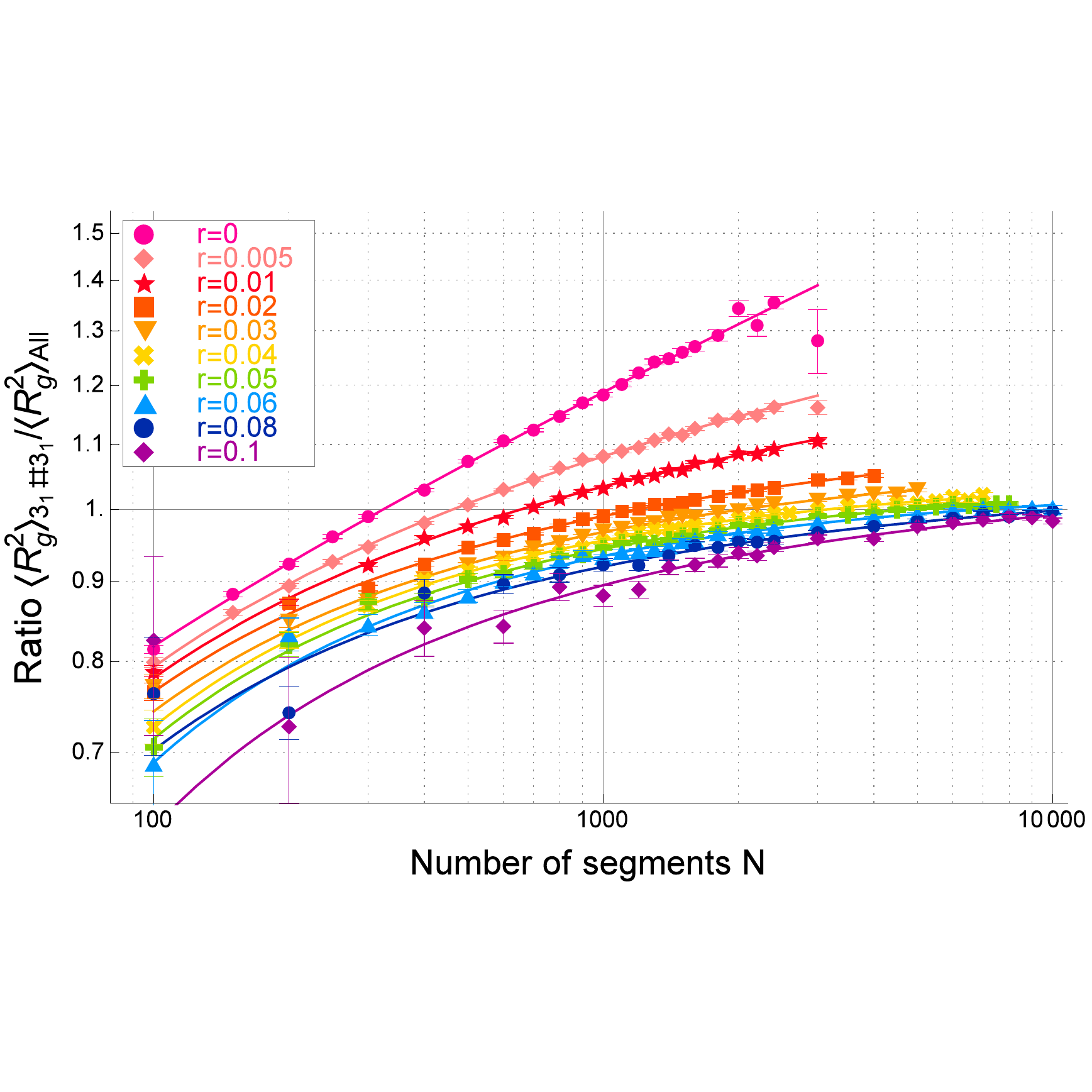}
\caption{Ratio of topological swelling for the composite knot $3_1 \# 3_1$ 
in cylindrical SAP with radius  $r_{\rm ex}$,  
$\langle R^2_g \rangle_{3_1 \# 3_1}/\langle R^2_{g} \rangle_{All}$,  
plotted against the number of segments $N$ in the double-logarithmic scale. }
  \label{fig:ratio3131}
\end{center}
\end{figure}

In order to express topological swelling graphically let us introduce the ratio of the mean-square radius of gyration for the cylindrical SAP with a knot $K$ to that of no topological constraint, $\langle R^2_{g} \rangle_{K}$/$\langle R^2_{g} \rangle_{All}$.  
We call it the ratio of topological swelling for knot $K$. 
The estimates of the ratio of topological swelling  are plotted against segment number $N$  in Figures \ref{fig:ratio01}, \ref{fig:ratio31} and  \ref{fig:ratio3131} for the trivial knot ($0_1$), the trefoil knot ($3_1$),  and composite knot $3_1 \# 3_1$, respectively, 
in the case of cylindrical SAP with different values of cylindrical radius $r_{\rm ex}$,  
in the double-logarithmic scale. 
They are also shown for the figure-eight knot ($4_1$) in Figure \ref{fig:ratio41} of section \ref{secA}. 
Here we remark that the composite knot $3_1 \# 3_1$ consists of two trefoil knots 
($3_1$).

In Figures \ref{fig:ratio01},  \ref{fig:ratio31}  and \ref{fig:ratio3131} if the excluded volume is small,  the ratio of topological swelling is larger than 1.0.  For instance, in the case of the trivial knot, the ratio of topological swelling is larger than 1.0 for all the values of the cylindrical radius, as shown in Figure  \ref{fig:ratio01}. For nontrivial knots, it is larger than 1.0 if the cylindrical radius is small enough and the number of segments is large enough.  In Figure \ref{fig:ratio31}  the ratio of topological swelling for radius $r_{\rm ex}=0.03$ increases with respect to the number of segments $N$ and becomes larger than 1.0 for  $N > 1, 000$ 
in the case of the trefoil knot ($3_1$).

The finite-size effect arising from a topological constraint is significant.
In the case of the trivial knot in Figure \ref{fig:ratio01} for the small values of radius such as $r_{\rm ex}=0.005$, 0.01 and 0.02  the ratio of topological swelling increases constantly with respect to the number $N$ of segments in the double-logarithmic scale at least within  the plotted range of  $N$ such as $N \le 4, 000$. However,  if we assume that the ratio of topological swelling would increase constantly even for asymptotically large values of $N$, then the scaling exponent of SAP with small positive values of radius $r_{\rm ex}$ should be larger than that of SAW, $\nu_{\rm SAW}$. Therefore, we suggest that the apparent constantly increasing behavior in the ratio of topological swelling with $r_{\rm ex} > 0$ in the double-logarithmic scale does not hold for asymptotically large values of $N$.  We expect that the gradient in the graph of the ratio of topological  swelling against  segment number $N$ should gradually decrease in the double-logarithmic scale 
as segment number $N$ increases and finally vanish for each nonzero value of radius $r_{\rm ex}$.  

In the experiment \cite{Takano12} the $g$-factors, $\langle R^2_{g} \rangle_{ring}/\langle R^2_{g} \rangle_{linear}$, are evaluated from the data of the gyration radii for ring polystyrenes in $\Theta$ solvents as $0.557 \le g \le 0.730$.  They are larger than 0.5, which is the $g$-factor for the RP with no topological constraint and RW. If we assume that the topologies of almost all the ring polystyrenes in the experiment are given by the trivial knot, it seems that the $g$-factors obtained in the experiment are consistent with the theoretical estimates of the ratio of topological swelling for zero radius ($r_{\rm ex}=0$) plotted in Figure \ref{fig:ratio01}.

%-----------------------------------------------------------------------
% Figure 2.  Radius of gyration 31 
%
%-----------------------------------------------------------------------

The finite-size effect plays an important role more explicitly for nontrivial knots, 
as shown in  Figures \ref{fig:ratio31}  and \ref{fig:ratio3131} for knots $3_1$  and $3_1 \# 3_1$, respectively.  In the case of $r_{\rm ex}=0.1$ the ratio of topological swelling increases and becomes close to the value of 1.0  only when the number of segments $N$ is very large such as $N=10, 000$. 
Here we remark that the characteristic length $N_{0_1}$ of the knotting probability  
is roughly estimated by 15, 000 in the case of $r_{\rm ex}=0.1$ for the cylindrical SAP \cite{UD2015}.  In Figure  \ref{fig:ratio31} the ratio for the trefoil knot $3_1$ increases very slowly: it is given by 0.8 at $N$=100, while it becomes close to 1.0 only at $N=10, 000$.

For the trivial knot the finite-size correction term vanishes when the cylindrical radius $r_{\rm ex}$ is large \cite{SAP02},  while for the nontrivial knots it becomes more significant as the cylindrical radius increases. It is clear in Figure \ref{fig:ratio01} that the graph of the ratio versus segment number $N$ has a more gradual rise, i.e., it increases slower, as the cylindrical radius gets  larger and finally becomes almost flat. However, in Figures \ref{fig:ratio31} and \ref{fig:ratio3131} the graphs of the ratio versus $N$ become more bending toward the lower direction in the small $N$ region and consequently have a sharper rise.  Therefore, the finite-size effect due to a topological constraint  is also significant for SAP even in the case of large excluded volume. 

We suggest that the slowly increasing behavior in the plots of  the ratio of topological swelling 
against segment number $N$,  which is particularly clear for non-trivial knots 
such as shown in Figures \ref{fig:ratio31} and \ref{fig:ratio3131},  
should be in agreement with the physical interpretation that we need to have a longer rope to tie a nontrivial knot if the thickness of the rope increases.  
Furthermore, we suggest that it corresponds to the slowly decaying behavior of the finite-size correction term in an appropriate fitting formula, which is applied to the plots of the ratio of topological swelling against segment number $N$.

We remark that topological swelling has been studied in Ref. \cite{SAP02} for the cylindrical SAP with several values of radius $r_{\rm ex}$. However,  the number of segments $N$ was limited such as $N \le10^3$ and some aspects of the finite-size effect have not been shown.  
In fact, the finite-size correction remains nontrivial even at $N=10^4$ 
for the cylindrical SAP with large values of  the radius such as $r_{\rm ex}=0.1$, as shown 
in Figures \ref{fig:ratio31} and \ref{fig:ratio3131}, while it was simply expected that  the ratio of topological swelling should approach 1.0 in any standard models of SAP as the number of segments $N$ goes to infinity,  
 if  the excluded volume is large enough \cite{SAP02}.

%-----------------------------------------------------------------------
% Figure 3 Radius of gyrations 3131
%
%-----------------------------------------------------------------------

%%%%%%%%%%%%
\subsection{Three-parameter formula for the ratio of topological swelling}

The finite-size effect plays an important role in the plots of the average size of SAP with a fixed knot against the number of segments $N$ plotted in Figures \ref{fig:ratio01}, \ref{fig:ratio31}  and  \ref{fig:ratio3131}, as shown  in the last subsection. It is effective almost through the whole range of the number $N$ of segments of SAP we investigated.  
It follows that  the plots do not completely show the asymptotic behavior even for $N=10^4$.  Thus, the   asymptotic expression is not necessarily appropriate to express them.

In order to express the strong finite-size effect we now introduce a three-parameter formula with scaling exponent as a fitting parameter. We also assume that  the finite-size correction is proportional to the inverse square root of $N$.  
\begin{equation}
\langle R^2_g \rangle_K/\langle R^2_{g} \rangle_{All}  = 
a_K \left( 1 + {\frac {b_K} {\sqrt{N}}}  \right) N^{2 \Delta \nu_K}   \, .  
\label{eq:ratioRK}
\end{equation}
Here,  $a_K$, $b_K$ and $\Delta \nu_K$ are the fitting parameters of  eq (\ref{eq:ratioRK}). 
We remark that the correction term corresponds to that of exponent $\Delta_1 \approx 0.5$ in the standard asymptotic expansion of the average size of SAW (see also \S 4.1).

We have applied eq (\ref{eq:ratioRK})  to the plots of the ratio of topological swelling versus $N$  
for several prime knots $0_1$, $3_1$ and $4_1$ together with 
some composite knots such as $3_1 \# 3_1$ and $3_1 \# 4_1$.    
Each of the fitted curves has a small $\chi^2$ value per degree of freedom (DF), so that they are good. The fitted curves are depicted in Figures  \ref{fig:ratio01}, \ref{fig:ratio31}  and \ref{fig:ratio3131} (for knot $4_1$, see Figure \ref{fig:ratio41} of section \ref{secA}).  We observe that they fit to the data points very well. 
The best estimates of knots $0_1$, $3_1$ and $3_1 \# 3_1$ are listed in Tables \ref{tab:ratio-01}, \ref{tab:ratio-31} and  \ref{tab:ratio-3131}, respectively. 
For knot $4_1$ they are given in Table \ref{tab:ratio-41} of section \ref{secA}.  

The finite-size correction term in eq (\ref{eq:ratioRK}) is appropriate to all the data shown in Figures  \ref{fig:ratio01}, \ref{fig:ratio31}   and \ref{fig:ratio3131}. 
In particular,  the monotonically and slowly increasing behavior 
of the ratio of topological swelling against segment number $N$,  
shown for the non-trivial knots in \ref{fig:ratio31},  
and \ref{fig:ratio3131},  
is very well described by the finite-size correction term of eq (\ref{eq:ratioRK}). 
Therefore,  the fitted curves have small values of the $\chi^2$ values per DF.

%%%%%%%%%%%%%%%%%%%%%%%%%%%%
%\input{table-ratio.tex}
%%%%%%%%%%%%%%%%%%%%%%%%%%%%  

% Mean-square radius of gyration
\begin{table}[htbp]
\begin{center} 
\begin{tabular}{c|cccc} \hline 
$r_{ex}$ & $a_K$ & $\Delta \nu_K $ & $b_K$ & $\chi^2/{\rm DF}$ \\  \hline 
0 & $0.456 \pm 0.017$ & $0.0682 \pm 0.0023$ & $2.39 \pm 0.19$ & $0.81$  \\ 
0.005 & $0.767 \pm 0.013$ & $0.0279 \pm 0.0011$ & $0.502 \pm 0.080$ & $0.69$  \\ 
0.01 & $0.878 \pm 0.012$ & $0.01586 \pm 0.00086$ & $0.155 \pm 0.061$ & $0.70$  \\ 
0.02 & $0.9462 \pm 0.0071$ & $0.00770 \pm 0.00047$ & $0.034 \pm 0.036$ & 
$0.50$  \\ 
0.03 & $0.9592 \pm 0.0063$ & $0.00532 \pm 0.00040$ & $0.048 \pm 0.032$ & 
$0.68$  \\ 
0.04 & $0.9786 \pm 0.0049$ & $0.00313 \pm 0.00030$ & $0.011 \pm 0.026$ & 
$0.67$  \\ 
0.05 & $0.9846 \pm 0.003$ & $0.00224 \pm 0.00018$ & $0.005 \pm 0.016$ & 
$0.35$  \\ 
0.06 & $0.9893 \pm 0.0032$ & $0.00161 \pm 0.00019$ & $-0.002 \pm 0.018$ & 
$0.45$  \\ 
0.08 & $0.9910 \pm 0.0025$ & $0.00111 \pm 0.00015$ & $0.007 \pm 0.014$ & 
$0.30$  \\ 
0.1 & $0.9932 \pm 0.002$ & $0.00078 \pm 0.00011$ & $0.007 \pm 0.011$ & 
$0.22$  \\ 
\hline 
\end{tabular}
\end{center} 
\caption{Best estimates of  the parameters in eq (\ref{eq:ratioRK}) 
for the ratio of topological swelling for the trivial knot ($0_1$),  
$\langle R^2_g \rangle_{0_1}/\langle R^2_{g} \rangle_{all}$.  }
\label{tab:ratio-01}
\end{table}

\begin{table}[htbp]
\begin{center} 
\begin{tabular}{c|cccc}  \hline 
$r_{ex}$ & $a_K$ & $\Delta \nu_K$ & $b_K$ & $\chi^2/{\rm DF}$ \\
 \hline 
0 & $0.524 \pm 0.016$ & $0.0615 \pm 0.0019$ & $-0.04 \pm 0.14$ & $0.96$  \\ 
0.005 & $0.794 \pm 0.017$ & $0.0269 \pm 0.0013$ & $-1.115 \pm 0.091$ & 
$1.15$  \\ 
0.01 & $0.911 \pm 0.020$ & $0.0148 \pm 0.0013$ & $-1.456 \pm 0.095$ & $1.76$  \\ 
0.02 & $1.000 \pm 0.017$ & $0.0054 \pm 0.0010$ & $-1.681 \pm 0.082$ & $1.95$  \\ 
0.03 & $1.0180 \pm 0.0085$ & $0.00262 \pm 0.00049$ & $-1.703 \pm 0.045$ & 
$0.66$  \\ 
0.04 & $1.0398 \pm 0.0080$ & $0.00053 \pm 0.00044$ & $-1.836 \pm 0.045$ & 
$0.74$  \\ 
0.05 & $1.035 \pm 0.010$ & $0.00029 \pm 0.00055$ & $-1.862 \pm 0.063$ & $1.3$  \\ 
0.06 & $1.024 \pm 0.010$ & $0.00049 \pm 0.00057$ & $-1.875 \pm 0.067$ & 
$1.26$  \\ 
0.08 & $1.006 \pm 0.010$ & $0.00101 \pm 0.00056$ & $-1.911 \pm 0.073$ & 
$0.85$  \\ 
0.1 & $0.990 \pm 0.017$ & $0.00160 \pm 0.00093$ & $-2.00 \pm 0.13$ & $1.65$  \\ 
\hline 
\end{tabular}
\end{center} 
\caption{Best estimates of the parameters in eq (\ref{eq:ratioRK}) for the ratio of 
topological swelling for the trefoil knot ($3_1$),  
$\langle R^2_g \rangle_{3_1}/\langle R^2_{g} \rangle_{all}$. }
\label{tab:ratio-31}.
\end{table}

\begin{table}[htbp]
\begin{center} 
\begin{tabular}{c|cccc} \hline
$r_{ex}$ & $a_K$ & $\Delta \nu_K$ & $b_K$ & $\chi^2/{\rm DF}$ \\
\hline 
0 & $0.496 \pm 0.023$ & $0.0654 \pm 0.0029$ & $-0.98 \pm 0.22$ & $1.30$  \\
0.005 & $0.823 \pm 0.026$ & $0.0253 \pm 0.0019$ & $-2.37 \pm 0.14$ & $1.22$  \\
0.01 & $0.917 \pm 0.027$ & $0.0148 \pm 0.0018$ & $-2.57 \pm 0.15$ & $1.28$  \\
0.02 & $1.014 \pm 0.025$ & $0.0050 \pm 0.0014$ & $-2.79 \pm 0.14$ & $1.24$  \\
0.03 & $1.038 \pm 0.024$ & $0.0020 \pm 0.0013$ & $-2.97 \pm 0.15$ & $1.18$  \\
0.04 & $1.076 \pm 0.020$ & $-0.0009 \pm 0.0010$ & $-3.19 \pm 0.14$ & $0.83$  \\
0.05 & $1.056 \pm 0.027$ & $-0.0004 \pm 0.0013$ & $-3.21 \pm 0.20$ & $1.11$  \\
0.06 & $1.068 \pm 0.026$ & $-0.0013 \pm 0.0012$ & $-3.47 \pm 0.20$ & $0.73$  \\
0.08 & $0.977 \pm 0.041$ & $0.0028 \pm 0.0021$ & $-3.0 \pm 0.4$ & $0.79$  \\
0.1 & $0.969 \pm 0.081$ & $0.0033 \pm 0.0042$ & $-3.73 \pm 0.80$ & $1.06$  \\
\hline 
\end{tabular}
\end{center} 
\caption{Best estimates of the parameters in eq (\ref{eq:ratioRK}) for the ratio of 
topological swelling for composite knot $3_1 \# 3_1$, 
$\langle R^2_g \rangle_{3_1 \# 3_1}/\langle R^2_{g} \rangle_{all}$.}
\label{tab:ratio-3131}.
\end{table}

In Figures  \ref{fig:ratio01}, \ref{fig:ratio31} and \ref{fig:ratio3131}  
we have observed that for the trivial knot the finite-size correction vanishes when the cylindrical radius $r_{\rm ex}$ is large,  while for the nontrivial knots it becomes more significant as the cylindrical radius increases. 
We now show it by observing how the best estimates of the parameter $b_K$ 
depends on the radius $r_{\rm ex}$.   
For the trivial knot, the coefficient $b_{0_1}$ in the finite-size correction term vanishes in the case of large values of cylindrical radius such as $r_{\rm ex} \ge 0.05$, as listed in Table \ref{tab:ratio-01}. However, for the nontrivial knots such as $3_1$ and $4_1$, the absolute value of the coefficient $b_{K}$ increases  for each knot as the cylindrical radius increases, as listed in Tables \ref{tab:ratio-31} 
(for knot $4_1$ see Table\ref{tab:ratio-41} of section \ref{secA}).

%%%%%%%%%%%%%%%%%%%%%%%%%%%%%%%%%%%%%%%%%%%%%%%%%%%%%%
\subsection{Characteristic length of the knotting probability and the equilibrium length of a knot}

We define the knotting probability of a knot $K$ for a model of  RP or SAP  
by the probability that  a given configuration of RP or SAP of $N$ segments in the model has the knot type $K$.  We denote it by $P_{K}(N)$. For a wide range of the number of segments $N$ 
we can approximate the knotting probability of a knot $K$ as a function of $N$ as 
\begin{equation} 
P_K(N)= C_K (N-\Delta N (K))^{m(K)} \exp(- \left( N -\Delta N(K) \right) /N_K ) \, ,  
\label{eq:knotPB}
\end{equation}
where $C_K$, $N_K$, $m(K)$ and $\Delta N(K)$ are fitting parameters \cite{JKTR,Deguchi-Tsurusaki1997}. 
It is shown for the cylindrical SAP that the estimates $N_K$ for many different knots $K$ are all approximately equal to that of the trivial knot: $N_K \approx N_{0_1}$ \cite{PLA2000,UD2015}. 
Thus,  we denote them simply by $N_0$ \cite{JKTR} and   
call it the characteristic length of the knotting probability. The estimates of $N_0$ 
for the cylindrical SAP with some values of radius $r_{\rm ex}$  
are  given in Table \ref{tab:equi-length-prime}.  We remark that the corrections 
$\Delta N(K)$ are smaller than the characteristic length $N_0$, and can be neglected 
for large $N$.

In Figure  \ref{fig:ratio31} we observe that  the ratio of topological swelling for knot $3_1$ 
reaches the value 1.0 when the number of segments $N$ is approximately equal to the  characteristic length $N_{0}$.   Here we recall that if the ratio of topological swelling for a knot $K$ 
becomes 1.0  in the RP or SAP with a knot $K$ of $N$ segments, then we call 
the number $N$ the equilibrium length of the knot $K$ with respect to the mean-square radius of gyration of the RP or SAP \cite{Rawdon08}.  
The estimates of the equilibrium length for simple prime knots are 
listed in Table \ref{tab:equi-length-prime}. 
In Figure \ref{fig:ratio3131} the ratio of topological swelling for 
composite knot $3_1 \# 3_1$ reaches the value 1.0 when segment number $N$ is approximately equal to twice the characteristic length $N_{0}$.

\begin{table}[htbp] 
\center 
\begin{tabular}{c|ccccc} 
\hline
$r_{\rm ex}$ & $N_0$ & $3_1$  & $4_1$ & $5_1$  & $5_2$  \\
\hline \hline 
0.0 &    $246 \pm 1$  & $197 \pm 8$ &  $291 \pm 21$ & $394 \pm 62$ & $387 \pm 37$ \\  
\hline 
0.01 &   $483 \pm 2$ & $353 \pm 8$ & $601 \pm 38$ & $869 \pm 92$ & $839 \pm 80$ \\  
\hline 
0.02 &   $818 \pm 3$  & $626 \pm 15$ & $1119 \pm 47$ & $1653 \pm 151$ & $1793 \pm 127$  \\  
\hline 
0.03 &   $1290 \pm 4$   & $1036 \pm 14$ & $2024 \pm 93$ & $2669 \pm 240$ & $2850 \pm 166$ \\  
\hline
0.04 &   $1966 \pm 5$  & $1610 \pm 15$ & $3298 \pm 108$ & $4106 \pm 324$ & 
$4820 \pm 320$  \\  
\hline
0.05 &   $2903 \pm 6$  & $2383 \pm 29$ & $5106 \pm 168$ & $6386 \pm 530$ & 
$10243 \pm 512$ \\  
\hline
\end{tabular} 
\caption{Equilibrium length $N_{\rm eq}$  for prime knots $3_1$, $4_1$, $5_1$ and $5_2$ and 
the characteristic length of the knotting probability $N_0$ versus radius $r_{\rm ex}$ of 
cylindrical segments. }
\label{tab:equi-length-prime}
\end{table}

Let us confirm numerically the above observation in Figs. \ref{fig:ratio31}  and \ref{fig:ratio3131}.         
We calculate the ratio of topological swelling, $\langle R^2_g \rangle_{K}/\langle R^2_{g} \rangle_{All}$, for a knot $K$ through the best estimates of the parameters $a_K$, $b_K$ and $\Delta \nu_K$ of the fitted curves given by eq (\ref{eq:ratioRK}) listed in Tables \ref{tab:ratio-31} and \ref{tab:ratio-3131}. 
For $r_{\rm ex} =0.02$ the ratio for the trefoil knot $3_1$ becomes 1.0 
at $N=626$, while that of the composite knot $3_1 \# 3_1$ becomes 1.0 
at $N=1244$.  
It is very close to twice the number 626, which is given by 1252.  
The characteristic length $N_0$ is approximately given by  820, where 
we have the ratio 
$N_{\rm eq}/N_0= 626/820 \approx 0.76$.   Thus, the average size of the SAP of knot $3_1$  
is equal to that of no topological constraint when segment number  $N$ 
is slightly smaller than $N_0$.

We now argue that the ratio of topological swelling 
for a composite knot  consisting of $n$ trefoil knots,  
$\langle R^2_g \rangle_{{3_1} \# \cdots \# {3_1}}/\langle R^2_g \rangle_{All}$,  
reaches or becomes greater than 1.0 around at $N = n N_0$.  
Hereafter we denote the composite knot  consisting of $n$ trefoil knots   
${3_1} \# \cdots \# {3_1}$  by $K_{comp}$. 

In terms of the knotting probabilities $P_K(N)$ the mean-square radius of gyration for SAP under no topological constraint, $\langle R^2_g \rangle_{All}$, is expressed as  the average of the mean-square radius of gyration for SAP with fixed knots $\langle R^2_g \rangle_{K}$ over all different knots.  We  classify all knots into such classes of knots consisting of $n$ prime knots for $n=1, 2, \ldots$. There are prime knots ($n=1$), the composite knots consisting of two prime knots ($n=2$), etc.  Here we assume that the trivial knot consists of zero prime knot ($n=0$).  We recall that for a given knot $K$ we denote the number of constituent prime knots by $|K|$. Then, we have  
\begin{equation}
\langle R^2_g \rangle_{All} = \sum_{n=0}^{\infty} \sum_{|K|=n}  \langle R^2_g \rangle_{K} P_K(N) \, . 
\label{eq:average}
\end{equation}
Here we assume that the symbol $\sum_{|K|=n}$ denotes the sum over all knots consisting of $n$ prime knots. 

It follows from eq (\ref{eq:knotPB}) that  
the knotting probability has the maximum value around at $N=m(K) N_0$.  
The knotting probability $P_K(N)$ becomes quite small for $N < m(K) N_0$ due to the power $\left( N/N_0 \right)^{m(K)}$ in eq  (\ref{eq:knotPB}) 
and decreases exponentially with respect to $N$ for $N > m(K) N_0$. 
Here we remark that the exponent $m(K)$ are close to the integers $n$ for composite knots 
consisting of $n$ prime knots \cite{JKTR,Deguchi-Tsurusaki1997}. Thus, if the number of segments $N$ is given by $N= n N_0$ for some integer $n$, the right-hand side of eq (\ref{eq:average}), i.e.,  the average of the mean-square radius of gyration over all knots,  is approximately given by the contribution from the knots consisting of $n$ prime knots. 

For simplicity, let us assume that simple knots consisting of $n$ prime knots have some similar values of the mean-square radius of gyration as the given composite knot $K_{comp}$ in common, or they are all smaller than $K_{comp}$ with respect to size.  We therefore have 
\begin{eqnarray}
\langle R^2_g \rangle_{All} & \approx &  \sum_{|K|=n}  \langle R^2_g \rangle_{K} P_K(N) 
\nonumber \\  
 &  \approx &  \langle R^2_g \rangle_{K_{comp}}  \sum_{|K|=n}   P_K(N)  <  \langle R^2_g \rangle_{K_{comp}} 
\qquad \mbox{for } \quad N = n N_0 \, \label{eq:approx}  
\end{eqnarray}
It follows from eq (\ref{eq:approx}) that  the ratio of topological swelling,  $\langle R^2_g \rangle_{K_{comp}} /\langle R^2_g \rangle_{All}$,  becomes greater than 1.0 around at $N = n N_0$. 

It seems that the evaluation $\langle R^2_g \rangle_{K_{comp}} /\langle R^2_g \rangle_{All}  > 1 $
at $N= n N_0$ is valid, although the assumption that  simple knots consisting of $n$ prime knots have similar average sizes does not necessarily always hold.    
For an illustration, let us consider the  $n=1$ case of eq (\ref{eq:approx}). 
It is easy to see that the derivation of eq (\ref{eq:approx}) is valid also for $n=1$.  
We observe in Table \ref{tab:equi-length-prime}  that 
the equilibrium length of knot $3_1$ is approximately equal to but 
slightly smaller than the characteristic length of 
the knotting probability $N_0$ for each value of radius $r_{\rm ex}$. 
Therefore,  the ratio of topological swelling is greater than 1.0 at $N=N_0$. 
We thus find that it is consistent with eq (\ref{eq:approx}) in the case of $n=1$.

%%%%%%%%%%%%%%%%%%%%%%%%%%%%%%
\subsection{Topological swelling among different knots}

%-----------------------------------------------------------------------
% Figure 4  Radius of gyration for various knots 
%
\begin{figure}[htbp]
\begin{center}
\includegraphics[width=
%12.5cm 
0.8\hsize]{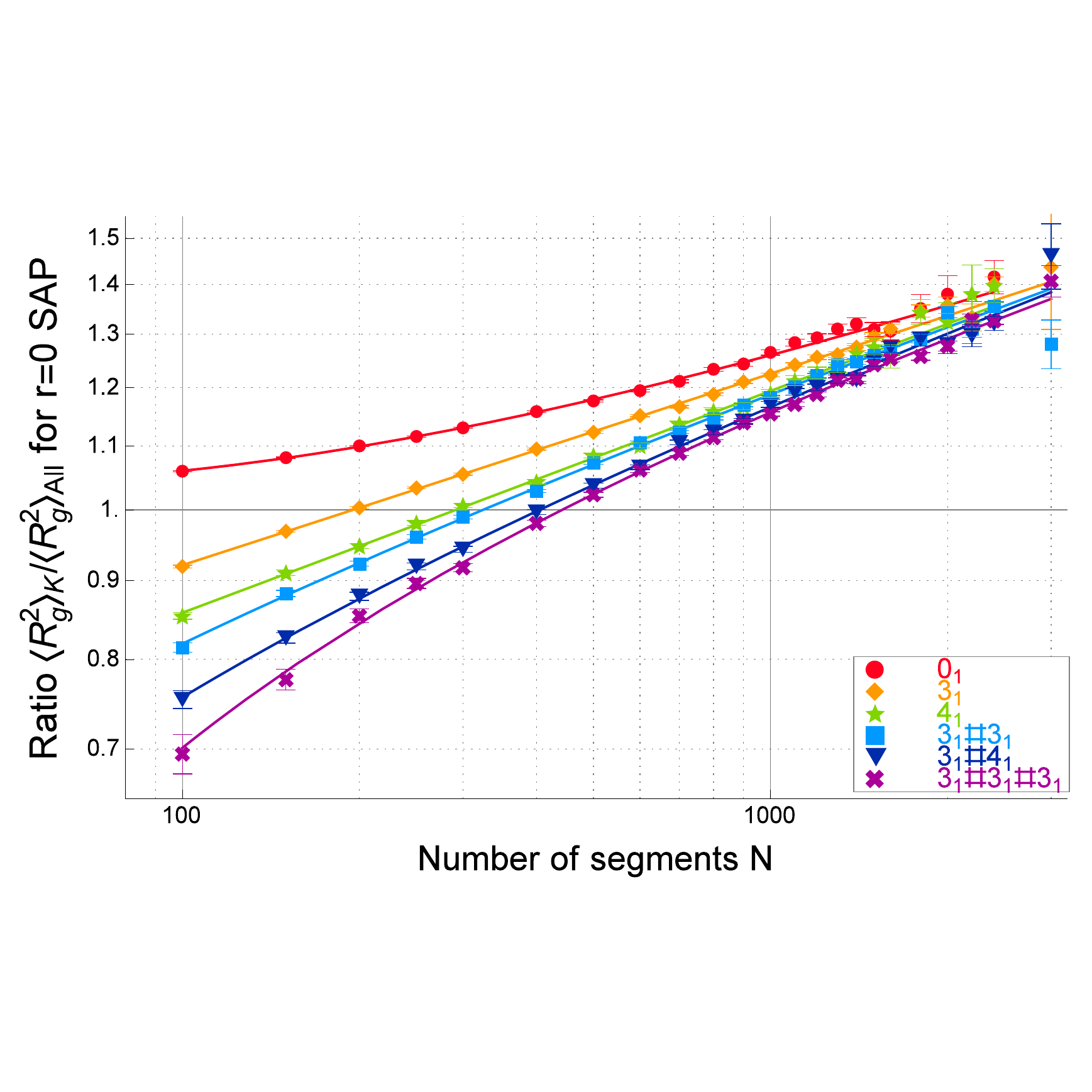}
\caption{Ratio of topological swelling for a knot $K$
in the equilateral RP,   $\langle R^2_g \rangle_{K}/\langle R^2_{g} \rangle_{All}$, is  plotted against the number of segments $N$ for various knots $K$ in the double-logarithmic scale. The plots for the trivial knot ($0_1$), the trefoil knot ($3_1$), the figure-eight knot ($4_1$), and three composite knots $3_1 \# 3_1$, $3_1 \# 4_1$ and $3_1 \# 3_1 \# 3_1$ are depicted by filled circles, diamonds, stars, squares, lower triangles and crosses, respectively, from $N$=100 to 3000. }
\label{fig:ratioK}
\end{center}
\end{figure}
%-----------------------------------------------------------------------

%%%%%%%%%%%%%%%%%%%%%%%%%%%%%%%%%%%%%%%%%%%%%%%%%%%%%%%%%%%%

We expect that the mean-square radius of gyration in  the cylindrical SAP with a fixed knot does not depend on the knot type if the number of segments $N$ is very large.  
If the knotted region of SAP with a fixed knot is localized along the polygonal chain, the difference in the average size among different knots should vanish as the number of segments $N$ goes to infinity.

In order to confirm it,  we plot the ratio of topological swelling against segment number $N$ for the trivial knot and some nontrivial knots.  
In Figure \ref{fig:ratioK} the ratio of topological swelling for the different knots in the equilatral RP,  
$\langle R^2_g \rangle_{K}/\langle R^2_{g} \rangle_{All}$, is plotted against the number of segments $N$. Here it is given by the cylindrical SAP in the case of zero thickness: $r_{\rm ex}=0$.  
The ratio becomes larger than 1.0 for each of the six knots if segment number $N$ is larger than 300 or 400. We observe that the ratios for the six knots become close to each other for large $N$ such as $N=2, 000$.  

For the equilateral random polygons the ratio of topological swelling for the trefoil knot ($3_1$) 
becomes 1.0  at $N=197$, while that of the composite knot $3_1 \# 3_1$  at $N=325$, which is slightly smaller than twice the number 197. Here the characteristic length $N_0$ is approximately given by 246, where we have the ratio $N_{\rm eq}/N_0 = 197/246 \approx 0.80$.  Thus, 
for $N=N_0$, the ratio of topological swelling for the trefoil knot ($3_1$) is larger than 1.0 as argued  in eq (\ref{eq:approx}).  For the figure-eight knot ($4_1$) the ratio of topological swelling  becomes 1.0 at $N=291$, which is slightly larger than $N_0$.

%***********************************************************************
% Section 4
% 
%***********************************************************************
%
\section{Scaling behavior of SAP with a fixed knot}
\subsection{Asymptotic expression of the mean-square radius of gyration}

We first review briefly some known results on the asymptotic behavior of the average size of polymer chains with excluded volume \cite{LGZJ,Nickel,LMS,GZJ,Clisby}. According to the standard renormalization group (RG) arguments \cite{Nickel} it is predicted that the mean-square radius of gyration $\langle R^2_g \rangle_{linear}$  of any real polymer chain should have the asymptotic behavior  
\begin{equation}
\langle R^2_g \rangle_{linear}  = A_{L} N^{2 \, \nu_{\rm SAW}}  
\left( 1 +  {b_{L}^{(1)} } {N^{-\Delta_1} }   + \cdots   \right) ,  
\label{eq:asymp-SAW}
\end{equation}
as $N$ increases infinitely \cite{LMS}. The critical exponents $\nu_{\rm SAW}$ and $\Delta_1$ are universal.  The estimates $\nu_{\rm SAW}=0.5880 \pm 0.0015$ and 
$\Delta_1 = 0.470 \pm 0.025$ were obtained by RG arguments via the $n=0$ limit of the $n$-vector field theory model \cite{LGZJ}, while the estimates $\nu_{\rm SAW}=0.5877 \pm 0.0006$ and $\Delta_1 = 0.56 \pm 0.03$ were obtained by the simulation of three-dimensional SAWs on lattice \cite{LMS}. 
 
We assume that the mean-square radius of gyration of the cylindrical SAP under no topological constraint, $\langle R^2_g \rangle_{All}$,  has the same asymptotic behavior as the mean-square 
radius of gyration of SAW, as $N$ goes to  infinity 
\begin{equation}
\langle R^2_g \rangle_{All}  = A_{R} N^{2 \, \nu_{\rm SAW}}  
\left( 1 +  {b_{R}^{(1)} } {N^{-\Delta_1} }  + \cdots  \right) . 
\label{eq:asympt}
\end{equation}
We employ eq (\ref{eq:asympt}) as a fitting formula with three parameters 
$A_R$, $b_{R}^{(1)}$ and $\Delta_1$.  
By putting $\nu_{\rm SAW}=0.588$ we apply eq (\ref{eq:asympt}) to the data of  $\langle R^2_g \rangle_{All}$ versus  $N$  for several different values of cylindrical radius $r_{\rm ex}$.  
The fitted curves given by eq (\ref{eq:asympt}) are good except for the case of $r_{\rm ex}=0$. The best estimates with $\chi^2$ values per DF  are listed in Table \ref{tab:asympt}. 
\begin{table}[htbp]
\begin{center} 
\begin{tabular}{c|cccc}
 \hline 
$r_{\rm ex}$ & $A_{R}$ & $\Delta_1$ & $b_{R}^{(1)}$ & $\chi^2/{\rm DF}$ \\
 \hline 
0 & $0.00200 \pm 0.00038$ & $0.1929 \pm 0.003$ & $43.1 \pm 8.1$ & $2.09$  \\
0.005 & $0.01843 \pm 0.00015$ & $0.2854 \pm 0.0033$ & $4.284 \pm 0.012$ & $1.04$  \\
0.01 & $0.024916 \pm 0.00009$ & $0.3273 \pm 0.0031$ & $3.029 \pm 0.020$ & $0.58$  \\
0.02 & $0.03267 \pm 0.00010$ & $0.3737 \pm 0.0062$ & $2.088 \pm 0.041$ & $1.64$  \\
0.03 & $0.037817 \pm 0.000078$ & $0.3903 \pm 0.0064$ & $1.566 \pm 0.035$ & $1.25$  \\
0.04 & $0.042129 \pm 0.000046$ & $0.4166 \pm 0.0053$ & $1.300 \pm 0.026$ & $0.65$  \\
0.05 & $0.045778 \pm 0.000039$ & $0.4305 \pm 0.0058$ & $1.061 \pm 0.024$ & $0.56$  \\
0.06 & $0.048918 \pm 0.00006$ & $0.4285 \pm 0.0099$ & $0.843 \pm 0.033$ & $1.02$  \\
0.08 & $0.054784 \pm 0.000054$ & $0.455 \pm 0.014$ & $0.599 \pm 0.034$ & $0.72$  \\
0.1 & $0.059973 \pm 0.000089$ & $0.426 \pm 0.032$ & $0.308 \pm 0.038$ & $1.35$  \\
\hline 
\end{tabular}
\end{center} 
\caption{Best estimates of the parameters of the asymptotic expansion (\ref{eq:asympt})
for the mean-square gyration radius of  the cylindrical SAP with radius $r_{\rm ex}$ under no topological constraint $\langle R^2_g \rangle_{All}$. } 
\label{tab:asympt}
\end{table}

We suggest that the estimate of $\Delta_1$ for the large radius case of $r_{\rm ex}=0.1$
 is consistent with that of SAW.  In Table \ref{tab:asympt} the estimate of $\Delta_1$ decreases as radius $r_{\rm ex}$ decreases. We have  $\Delta_1=0.426 \pm 0.032$ at $r_{\rm ex}=0.10$, which is close to the estimate $\Delta_1= 0.47 \pm 0.03$ derived through the $n=0$ limit of the field theory.  

We have applied the asymptotic expression of eq (\ref{eq:asympt}) also 
to the data  of the mean-square radius of gyration for the cylindrical SAP with a fixed knot $K$ denoted by  $\langle R^2_{g} \rangle_{K}$ versus segment number $N$.  For the trivial knot, we have good fitted curves.  For nontrivial knots,  however, we do not always have good fitted curves 
for all values of radius $r_{\rm ex}$.

For lattice SAP the asymptotic expression of eq (\ref{eq:asympt}) has been applied to the data of the mean-square radius of gyration under a topological constraint \cite{Marcone07}. It was suggested that the estimate $\Delta_1=1/4$ should be the most favorable to the data, although other values such as $\Delta_1=1/2$ are not completely denied. Thus, the correction term 
in the asymptotic expansion has not been  numerically determined, yet.

%%%%%%%%%%%%%%%%%%%%%%%%%%%%%%%%%%
\subsection{Fitting formula with an effective scaling exponent}

We now introduce the effective scaling exponent through curves fitted to the data of the mean-square radius of gyration for the cylindrical SAP with a fixed knot $K$, $\langle R^2_{g} \rangle_K$,  versus segment number $N$.   
We consider a three-parameter formula with a scaling exponent to be fitted where the finite-size correction is proportional to the inverse of the square root of segment number $N$ 
\begin{equation}
\langle R^2_{g} \rangle_K = A_K\left( 1 + {\frac {B_K} {\sqrt{N}}}  \right) N^{2 \nu_K} \, . 
\label{eq:RK}
\end{equation}
Here, the exponent $\nu_K$, the amplitude $A_K$ and the coefficient $B_K$ are fitting parameters.  We call the parameter $\nu_K$  the effective scaling exponent of knot $K$.

By applying eq (\ref{eq:RK}) to the data of the mean-square radius of gyration for cylindrical SAP
 with a fixed knot versus segment number $N$, we observe  that fitted curves are appropriate to  the plots of all the knot types and for all the ten different values of cylindrical radius $r_{\rm ex}$ from 0 to 0.1, as shown  in the captions.  The $\chi^2$ values per DF are small such as less than 2.0 for almost all fitted curves. For instance, 
the best estimates of the parameters of eq (\ref{eq:RK}) 
with the $\chi^2$ values  for knots $0_1$, $3_1$, and $4_1$ 
are listed in Tables \ref{tab:RK034}  of section \ref{secB}.  
For other seven knots such as $5_1$, $5_2$, $3_1\# 3_1$, $3_1 \# 4_1$,  $3_1\# 3_1 \# 3_1$, $3_1 \# 3_1 \# 4_1$, and $3_1 \# 4_1 \# 4_1$  the best estimates of the parameters of eq (\ref{eq:RK}) 
are also listed in Table \ref{tab:RK5152} - \ref{tab:RK314141} of section \ref{secB}.

The formula of eq (\ref{eq:RK}) corresponds to that of eq (\ref{eq:ratioRK}) if we have good fitted curves to the plots of the mean-square radius of gyration for the cylindrical SAP under no topological constraint, $\langle R^2_{g} \rangle_{All}$, against segment number $N$. The $\chi^2$ values are not large,  as  given together with the best estimates of the paraeters in Table \ref{tab:gyr-all} of section \ref{secC}.

\subsection{Finite-size correction term}

It is not trivial to choose an appropriate correction term $1/N^{\Delta}$,  as suggested in Ref. \cite{Madras-Slade}. We therefore simply put $\Delta=1/2$ in eq (\ref{eq:RK}).
Here we remark that in the perturbative approach to the excluded-volume effect the mean-square radius of gyration of a polymer is expanded in terms of the inverse square root of segment number $N$ \cite{Yamakawa}.

If we choose other powers of $N$ as the finite-size correction term in eq (\ref{eq:RK}) such as the inverse of  $N$, the  $\chi^2$ values per DF are not small  and hence the fitted curves derived  are not good.  
We have also applied the formula of eq (\ref{eq:RK}) with no finite-size correction term, i.e. we put $B_K=0$. The fitted curves are not good with respect to the $\chi^2$ values.  
By applying  eq (\ref{eq:RK}) with $B_K=0$  
to the data points of the mean-square  gyration radius for the equilateral RP with the trivial knot 
($0_1$) evaluated in the plot range of  $N \le 2400$, we have obtained the estimate of 
%of effective scaling exponent 
$\nu_0=0.53$, where the $\chi^2$ values per DF are rather large such as 18.

In the experiment it is shown that the mean-square radii of gyration of ring polystyrenes in $\Theta$ solvents are scaled with an enhanced exponent $0.53$, where the formula of eq (\ref{eq:RK}) with no finite-size correction term  is applied to the experimental data \cite{Takano12}. Thus, the estimate of the scaling exponent in the experiment coincides with that of the present research for the trivial knot if we apply eq  (\ref{eq:RK}) with no correction term (i.e., $B_K=0$). We conclude that the experimental estimate of the scaling exponent  \cite{Takano12} is not in contradiction with the theoretical estimate of the trivial knot evaluated in the present paper.

%%%%%%%%%%%%%%%%%%%%%%%%%%%%%%%%%%%%%%%%%%%%%
% section 5 
%
\section{Scaling exponent of the RP with a knot} 

\subsection{ Continuous change in effective exponent 
%$\nu_K$ 
with respect to radius} 
%$r_{\rm ex}$ }

\begin{figure}[htbp]
\begin{center}
\includegraphics[width=
%12.5cm
0.8 \hsize]{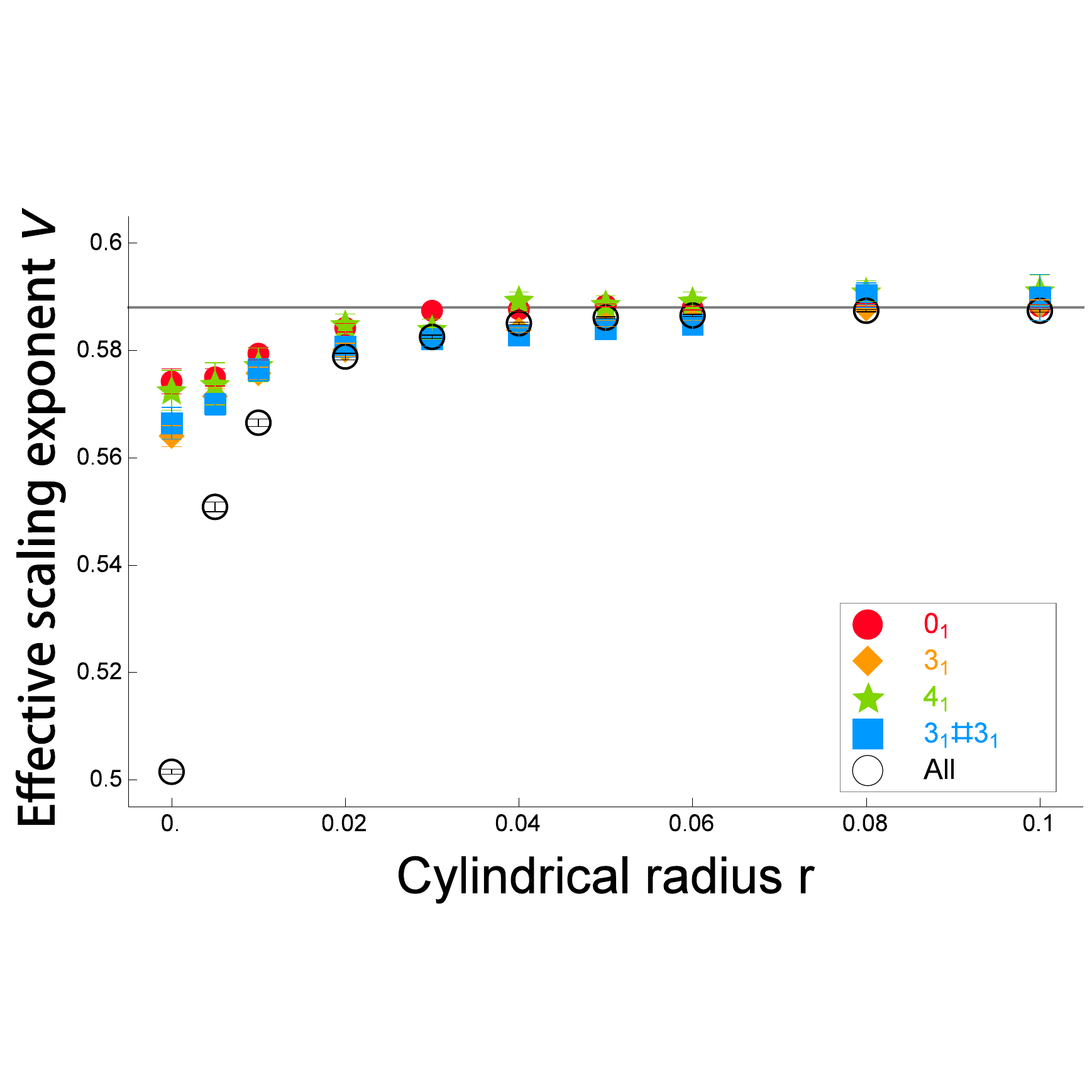}
\caption{
Best estimates of effective scaling exponent $\nu_K$ defined by eq (\ref{eq:RK})
for the mean-square radius of gyration of the cylindrical SAP with a knot $K$ versus radius $r_{\rm ex}$ of cylindrical segments in the case of  the trivial, trefoil, figure-eight knots and composite knot $3_1 \# 3_1$ are depicted by filled crosses,  circles, diamonds, and squares, respectively. Those of SAP under no topological constraint $\nu_{All}$ 
(All)  are depicted by open circles. } \label{fig:scaling-exponent}
\end{center}
\end{figure}

In Figure \ref{fig:scaling-exponent} the best estimates of the effective scaling exponent $\nu_K$ 
for the cylindrical SAP with a knot $K$ are plotted against cylindrical radius $r_{\rm ex}$  for the four knots: $0_1$, $3_1$, $4_1$ and $3_1 \# 3_1$. We observe that for each knot $K$ the estimate of $\nu_K$ decreases very slowly as cylindrical radius $r_{\rm ex}$ decreases and approaches zero. In particular, the change of the effective scaling exponent $\nu_K$ near the origin at $r_{\rm ex}=0$ is rather small. On the other hand, the effective scaling exponent of the cylindrical SAP under no topological constraint, $\nu_{All}$,  abruptly decreases near the origin at $r_{\rm ex}=0$ as cylindrical radius $r_{\rm ex}$ decreases and approaches zero. We have $\nu_{All}=0.5$ at $r_{\rm ex}=0$, while we have $\nu_{All}=0.55$ for a small nonzero radius of $r_{\rm ex}=0.005$.

We thus propose a conjecture that  the effective scaling exponent $\nu_K$ of the cylindrical SAP with 
radius $r_{\rm ex}$ of any given knot $K$ should be continuous as a function of cylindrical radius $r_{\rm ex}$ particularly near the origin of $r_{\rm ex}=0$.

%-----------------------------------------------------------------------
% Figure 5.
%

Even in the case of zero radius (i.e., $r_{\rm ex}=0$), the effective scaling exponent $\nu_K$ of a knot $K$ is definitely larger than  0.5.  We observe such enhancement of the effective scaling exponent 
$\nu_K$  clearly in Figure \ref{fig:scaling-exponent}.  Here, the scaling exponent of RW and RP is given by 0.5,  and we denote it by $\nu_{\rm RW}$. 
The estimates of  $\nu_K$ at  $r_{\rm ex}=0$ are distinct from the value of $\nu_{\rm RW}$ with respect to error bars which are given by the standard deviations. 
We have thus described topological swelling for random knots in terms of the effective scaling exponent $\nu_K$.

The effective scaling exponent $\nu_K$ for the equilateral RP with a knot $K$ is distinctly  smaller than that of SAW, $\nu_{\rm SAW}=0.588$, with respect to errors, as shown in Figure \ref{fig:scaling-exponent}.  The estimates $\nu_K$ for some knots are evaluated in the plot range of $N$ with the upper limit of $N= 3, 000$.  
For zero radius ($r_{\rm ex}=0$)  the estimates $\nu_K$ of the trivial and the trefoil knots are given by $\nu_{0_1}= 0.574 \pm 0.002$ and  $\nu_{3_1}= 0.563 \pm 0.002$, respectively.  They are listed in Table \ref{tab:RK034} of section \ref{secB}.

%
%%%%%%%%%%%%%%%%%%%%%%%%%%%%%%%%%%%%%%%%%%%%%%%%%%%%%%%%%%%%%%%%%%%%

\subsection{Enhancement in effective scaling exponent of the RP with a knot}

We now argue that the effective scaling exponent $\nu_K$ of  
the equilateral RP with a knot $K$ approaches the scaling exponent of SAW: $\nu_{\rm SAW}=0.588$, if the upper limit in the plot range of segment number  $N$ goes to infinity.  
Hereafter, we call  the upper limit in the plot range of  $N$ the maximum segment number $N$ 
or the maximum of $N$, in brief.

%-----------------------------------------------------------------------
% Figure 6.
%
\begin{figure}[htbp]
\begin{center}
\includegraphics[width=
%10.5cm
0.7 \hsize
]{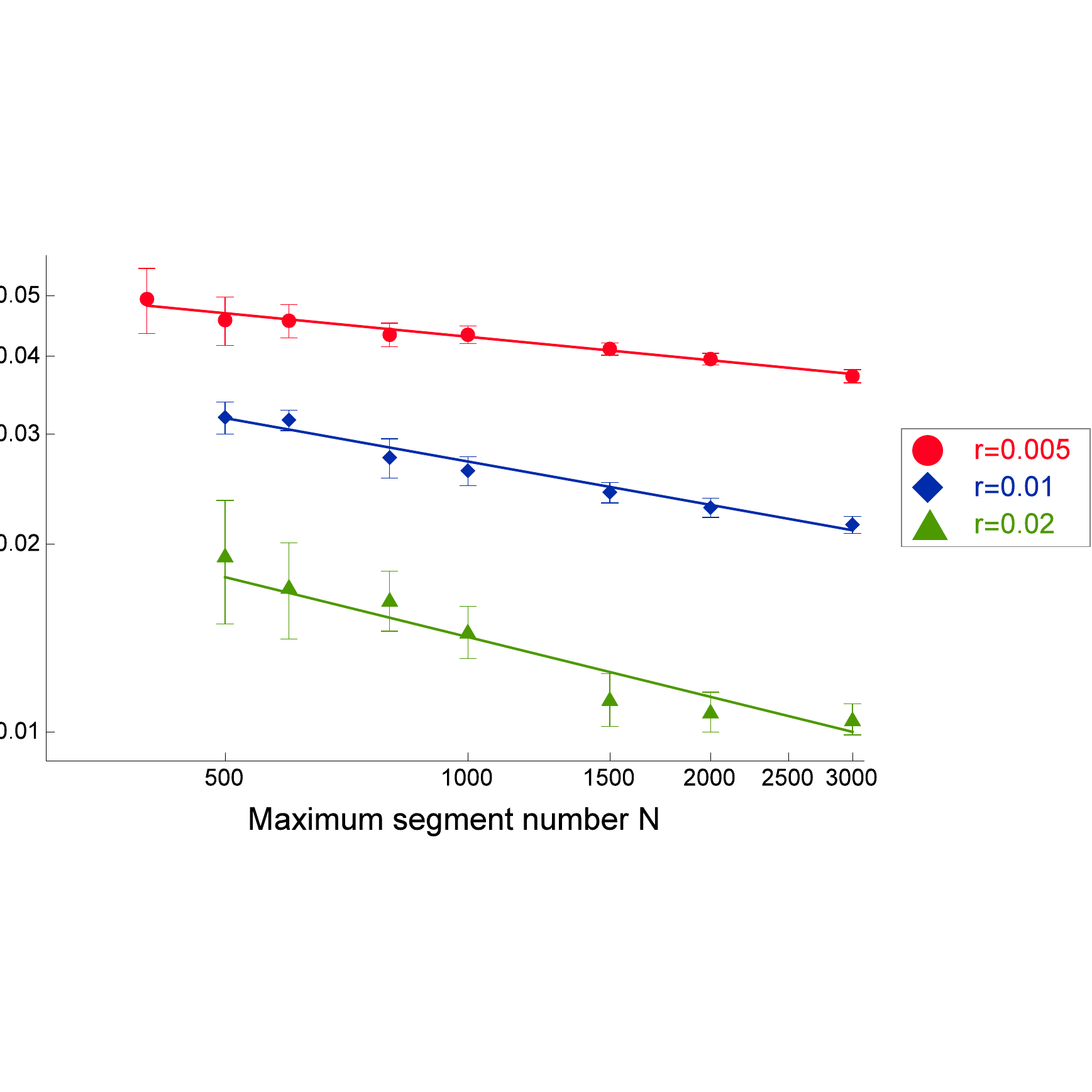}
\caption{
Difference of effective scaling exponent $\nu_{All}$ to scaling exponent of SAW 
$\nu_{\rm SAW}$:   $\nu_{\rm SAW}-\nu_{All}$ versus maximum segment number $N$ 
(i.e., the upper limit of the plot range of $N$) in the double-logarithmic scale. 
The plots of $r_{\rm ex}$= 0.005, 0.01 and 0.02 are depicted by 
filled circles, diamonds and upper triangles, respectively.  
Here we put $\nu_{\rm SAW}=0.588$. 
} \label{fig:difference_exponent}
\end{center}
\end{figure}

Let us assume that for any small value of the excluded volume the effective scaling exponent of SAW approaches the value of $\nu_{\rm SAW}$ if the maximum segment number $N$ becomes very large. For instance, the effective exponent $\nu_{All}$ defined by eq (\ref{eq:RK}) for the cylindrical SAP under no topological constraint (i.e.,  $K=All$)  with radius $r_{\rm ex}=0.01$ is given by $\nu_{All}=0.5665 \pm 0.0007$ evaluated in the plot range from  $N=100$ to  $N=3, 000$. However, we expect that if we increase the maximum segment number $N$ to a very large number for the cylindrical SAP with $r_{\rm ex}=0.01$, then the effective scaling exponent $\nu_{All}$ should increase and become much closer to the value of $\nu_{\rm SAW}=0.588$. Here we remark that the estimates  $\nu_{All}$ for other values of cylindrical radius $r_{\rm ex}$ are listed in Table \ref{tab:gyr-all} of section  \ref{secC}.

In Figure  \ref{fig:difference_exponent}, the difference between the effective scaling exponent $\nu_{All}$ and the scaling exponent of SAW $\nu_{\rm SAW}$  is plotted against the upper limit in the plot range of segment number $N$ for the cylindrical SAP 
in the case of radius $r_{\rm ex}=0.005$, 0.01 and 0.02, respectively,  in the double-logarithmic scale. 
We observe that the plots can be approximated by linear lines in all the three cases. 
We also observe in Figure \ref{fig:difference_exponent}  that 
the absolute value of the gradient of the fitted line increases 
as the cylindrical radius becomes larger: the more the excluded volume is, 
the faster $\nu_{All}$ approaches $\nu_{\rm SAW}$.

In the case of $r_{\rm ex}=0.005$  the fitted line is the best among the three cases of values of radius $r_{\rm ex}$.  We suggest that the effective scaling exponent $\nu_{All}$ approaches the scaling exponent of SAW where  the difference of exponents: $\nu_{\rm SAW}-\nu_{All}$ is roughly approximated by  
the inverse power of $N$ with a small exponent such as 0.125 with a constant $c_{All}$: 
\begin{equation} 
\nu_{All} = \nu_{\rm SAW} - c_{All} N^{-0.125}  \, . 
\end{equation}
Thus, it approaches $\nu_{\rm SAW}$ very slowly.  For instance,   
the difference between $\nu_{\rm SAW}$ and $\nu_{All}$ is reduced to 0.01, i.e.,  $\nu_{All}=0.587$,  if the maximum segment number $N$ is very large such as $N=10^8$. 
Here we remark that the power-law decay is considered as an approximate 
expression, since we have confirmed it only over a decade from $N=300$ to 3, 000.

For any other topological condition $K$ than no topological constraint  ($All$) 
such as being topologically equivalent to a knot $K$, it is not practically easy to show numerically 
how  the effective scaling exponent $\nu_K$ for the cylindrical SAP with a small nonzero radius $r_{\rm ex}$ approaches $\nu_{\rm SAW}$ as the maximum segment number $N$ increases.

However, from Figures  \ref{fig:scaling-exponent} and \ref{fig:difference_exponent} 
we argue that the effective scaling exponent $\nu_K$ of the cylindrical SAP with any given knot $K$ approaches the scaling exponent of SAW $\nu_{\rm SAW}$ if the maximum segment number  $N$ becomes very large.  The way it approaches  $\nu_{\rm SAW}$ can be very slow 
such as that it approaches 0.587 (i.e. the difference from $\nu_{\rm SAW}$ is given by 0.01) when the maximum of $N$ is as large as $N=10^8$ or larger. 

In Figure \ref{fig:scaling-exponent} we observe that the estimates of effective scaling exponents $\nu_{K}$ for several knots $K$ are all larger than the estimate of $\nu_{All}$  for almost any value of radius $r_{\rm ex}$. We thus expect that the effective exponents $\nu_{K}$ are larger than $\nu_{All}$ for any value of cylindrical radius $r_{\rm ex}$ even if  the maximum of $N$ becomes extremely large.  
Thus,  if the maximum segment number $N$ becomes very large,  
the effective scaling exponent $\nu_K$ for the cylindrical SAP of a knot $K$ 
even with a small radius $r_{\rm ex}$ approaches the value of $\nu_{\rm SAW}$ very closely, since $\nu_{All}$ increases to $\nu_{\rm SAW}$  and $\nu_{K}$ is assumed 
to be larger than $\nu_{All}$. 

We now argue enhancement of the effective scaling exponent $\nu_K$ of the RP with a knot $K$.  
It follows from the continuity conjecture of effective exponents $\nu_K$ 
with respect to radius $r_{\rm ex}$ shown in section 5.1 
that the effective scaling exponent $\nu_K$ of  
the equilateral RP with a knot $K$ approaches $\nu_{\rm SAW}$ if the upper limit in the plot range of $N$ (i.e., the maximum of $N$) goes to infinity. We expect that the difference $\nu_{\rm SAW}-\nu_{K}$ is smaller than 0.01 if the maximum of $N$ is given by $10^8$. If the estimate of $\nu_{K}$ at $r_{\rm ex}=0.005$ becomes equal to $\nu_{\rm SAW}$, the estimate of $\nu_{K}$ at $r_{\rm ex}=0.0$ becomes  also equal to it, since the effective exponent $\nu_{K}$ is assumed to be continuous with respect to radius $r_{\rm ex}$.

We give a conclusion to the above argument:  The effective scaling exponent $\nu_K$ of the RP with a fixed knot $K$ should approach the value of $\nu_{\rm SAW}$ if the upper limit in the plot range  of segment number $N$ is very large  such as $N=10^8$. However, the topological finite-size effect is very strong, so that it is effectively smaller than the value of $\nu_{\rm SAW}$ if segment number $N$ is less than $10^4$.

%%%%%%%%%%%%%%%%%%%%%%%%%%%%%%%%%%%%%%%%%%%%%%%%%%%%%%%%%%
%
% Sec 6
%
\section{Additivity of equilibrium lengths for composite knots}  

\subsection{Equilibrium lengths of composite knots }

Associated with topological swelling we propose the following conjecture:   
If the equilibrium lengths for knot $K_j$ are given by $N_{{\rm eq}}(K_j) $ for $j=1$ and $2$, 
respectively, then the equilibrium length for the composite knot $K_1 \# K_2$, 
denoted by $N_{\rm eq}(K_1 \# K_2)$,  is given by the sum 
\begin{equation}
N_{\rm eq}(K_1 \# K_2) = N_{\rm eq}(K_1) + N_{\rm eq}(K_2) .  
\end{equation} 

\begin{table}[htbp] 
\center 
\begin{tabular}{c|ccccc} 
\hline
$r_{\rm ex}$ & 
$3_1 \# 3_1$  & $3_1\# 4_1$ 
&  $3_1 \# 3_1 \# 3_1$  & $3_1 \# 3_1\# 4_1$ & $3_1 \# 4_1\# 4_1$   \\
\hline \hline 
0.0 &  
$325 \pm 19$ & $405 \pm 33$ & $442 \pm 29$ & $528 \pm 51$ & $621 \pm 76$ \\  
Fraction & 
$0.82 \pm 0.05$ & $0.83 \pm 0.07$ & $0.75 \pm 0.05$ &  $0.77 \pm 0.07$ & $0.80 \pm 0.10$  \\  
\hline 
0.01 &  
$654 \pm 30$ & $900 \pm 57$  & $943 \pm 72$ & $1167 \pm 97$ & $1426 \pm 251$ \\  
Fraction   & 
$0.93 \pm 0.04$ & $0.94 \pm 0.06$  & 
$0.89 \pm 0.07$  & 
$0.89 \pm 0.07 $  & 
$0.92 \pm 0.16 $  \\  
\hline
0.02 &   
$1244 \pm 51$ & $1740 \pm 115$ & $1743 \pm 98$ & $2272 \pm 179$ & $2717 \pm 587$ \\  
Fraction   &
$0.99 \pm 0.04$ & 
$1.00 \pm 0.07$  & $0.93 \pm 0.05$  & $0.96 \pm 0.08 $  & $0.95 \pm 0.20 $  \\  
\hline 
0.03 &   
$2081 \pm 71$ & $2696 \pm 167$ & $2907 \pm 230$ & $3951 \pm 450$ & $4468 \pm 894$ \\  
Fraction   &
$1.00 \pm 0.03$ & $0.88 \pm 0.05 $  & $0.94 \pm 0.07 $  & $0.96 \pm 0.11$  
& $0.88 \pm 0.18 $  \\  
\hline
0.04 & 
$3217 \pm 85$ & $4430 \pm 323$ & $4655 \pm 317$ & $5821 \pm 687$ & $7757 \pm 2187$ \\  
Fraction   &
$1.00 \pm 0.03$ & $0.90 \pm 0.07$  & $0.96 \pm 0.07$  & $0.89 \pm 0.11 $  
& $0.95 \pm 0.27 $  \\  
\hline
0.05 & 
$4869 \pm 168$ & $6452 \pm 413$ & $7053 \pm 564$ & $10203 \pm 1245$ & $14339 \pm 2882$ \\  
Fraction   &
$1.00 \pm 0.03$ & $0.86 \pm 0.06$  & $0.99 \pm 0.08 $  & $1.03 \pm 0.13$  
& $1.14 \pm 0.23 $  \\  
\hline
\end{tabular} 
\caption{ Equilibrium lengths for  composite knots of knots $3_1$ and $4_1$. Fraction shows the ratio of the equilibrium length of a composite knot to the sum of the equilibrium lengths over all the constituent prime knots together with errors.}
\label{tab:equi-length}
\end{table}

%%%%%%%%%%%%%%%%%%%%%%%%%%%%%%%%%

For an illustration, let us confirm the conjecture for composite knot $3_1 \# 4_1$  numerically 
by making use of the best estimates for the parameters of the formula of eq (\ref{eq:ratioRK}). 
For $r_{\rm ex} =0.02$ the ratio of topological swelling for the trefoil knot ($3_1$) 
becomes 1.0 at $N=626$ and for the figure-eight  knot ($4_1)$ at $N=1119$, while that of composite knot $3_1 \# 4_1$ at $N=1740$. It is almost equal to the sum of the numbers $N_1=626$ and $N_2=1119$, which is given by 1745. Thus, the ratio    
of the equilibrium length of  the composite knot $3_1 \# 4_1$ 
to the sum of those of knots $3_1$ and $4_1$ is given by 1.0 with respect to errors.

Several examples support numerically the additivity conjecture. 
In Table  \ref{tab:equi-length} the estimates of the equilibrium length for five 
composite knots are listed for several different values of radius $r_{\rm ex}$ of cylindrical segments.   In each row of a given value of radius $r_{\rm ex}$ the row of ``Fraction'' shows the ratio  of the equilibrium length of a given composite knot, $N_{\rm eq}({\sum_{j} \# K_{j}})$, which we denote simply by $N_{12}$,    
to the sum of the equilibrium lengths of all constituent prime knots,  
 $\sum_{j} N_{\rm eq}({K_j})$, which we denote simply by $N_1 +N_2$.  
The fraction is denoted by  $N_{12}/(N_1+N_2)$. 
They are close to 1.0  with respect to errors  at least for  SAP.

%%%%%%%%%%%%%%%%%%%%%%%
\subsection{Additivity compatible with the local  knot conjecture}
 
Let us now argue that the additivity of equilibrium lengths for composite knots is compatible and even consistent with the local knot conjecture.  The conjecture is given as follows. In an ensemble of  such RP or SAP with a fixed prime knot the majority of them have such knotted regions that are localized along the polygonal chains \cite{Orlandini1998,Katritch00,Marcone05}.  
In a given SAP with a prime knot $K$, if the knotted region is localized as in the local knot conjecture,  we can take a subchain which corresponds to the knotted region. We call the subchain a locally knotted SAW with the knot $K$, which consists of SAW with the local knot $K$ on it.  
We also assume that if we connect the two ends of a locally knotted  SAW with a knot $K$, we have SAP with the knot $K$. Under the local knot conjecture we expect that the majority of SAPs with a knot $K$ have locally knotted SAW with the knot $K$ as subchains of the SAP.

Recall that if the equilibrium length for a knot $K$  is given by $N$, the mean-square radius of gyration of SAP with the fixed knot $K$ is equal to that of SAP under no topological constraint.  
Here, we physically interpret that topological entropic repulsions are balanced with the complexity of the knot $K$ in the SAP with the knot $K$ of $N$ segments.  
Thus, also  in $N$-step locally knotted SAWs with the knot $K$, we say that  
topological entropic repulsions are balanced with the complexity of the knot $K$
if the step number $N$ is equal to the equilibrium length for the knot $K$.

Suppose that in a SAP with a prime knot $K_1$ of $N_1$ segments  
topological entropic repulsions are balanced with the complexity of the knot $K_1$ and  
also that in  a SAP with a prime knot $K_2$ of $N_2$ segments 
topological entropic repulsions are balanced with the complexity of the knot $K_2$.  
According to the local knot conjecture, in the locally knotted $N_j$-step SAWs with the knot $K_j$ topological entropic repulsions are balanced with the complexity of the knot $K_j$ for $j=1$ and 2, respectively.  
We then suggest that the topological entropic repulsions in the SAP of $N_1 + N_2$ segments  with the composite knot $K_1 \# K_2$ should be balanced with the complexity of the knot $K_1 \# K_2$. 
Here, we expect that in the composite SAW 
of $N_1 + N_2$ steps with the composite knot $K_1 \# K_2$ there are 
two locally knotted regions which correspond to the two prime knots $K_1$ and $K_2$, respectively.  Since in each of  the $N_j$-step locally knotted SAWs ($j=1, 2$) 
the topological entropic repulsions are balanced with the knot complexity of the knot $K_j$,  
the whole topological entropic repulsions in the composite SAW of $N_1 + N_2$ steps  with the knot $K_1 \# K_2$ should be balanced with the complexity of the knot $K_1 \# K_2$. 
Under the local knot conjecture it follows that the equilibrium length of   
the total SAP of $N_1+N_2$ segments with the knot $K_1 \# K_2$ is given by 
the sum $N_1 + N_2$. 

\subsection{Possibility of knots in RP being less localized than in SAP}
We now argue that the local knot conjecture holds for SAP well, while it holds for RP less than for  SAP.  In Table \ref{tab:equi-length} we observe that 
the fractions  $N_{12}/(N_1 + N_2)$  are given from 0.9 to 1.0 for SAP, while it is about 0.8 for RP. 
They are smaller in RP than in SAP for the five composite knots. 
Here we expect that if knots are truly localized, then the fraction   $N_{12}/(N_1 + N_2)$ should be given by 1.0.  We therefore suggest that the knotted region in the RP with a fixed knot  is less localized than in the SAP with a fixed knot.   

We recall that 
in Figure \ref{fig:ratioK} the ratio of topological swelling in the equilateral RP,  
$\langle R^2_g \rangle_{K}/\langle R^2_{g} \rangle_{All}$, is plotted against the number of segments $N$  for the six knots.

We suggest that topological entropic repulsions are stronger in knotted RP than in knotted SAP. 
If knotted regions are less localized in knotted RP than in knotted SAP, they should be more entangled in knotted RP and hence the topological entropic repulsions among segments should be stronger in knotted RP than in knotted SAP.  

It may explain the reason why the fractions $N_{12}/(N_1 + N_2)$ are smaller in knotted RP than those of knotted SAP.  
If topological entropic repulsions among segments are stronger in knotted RP,  then the equilibrium length of a knot is smaller  in RP than in SAP with respect to the characteristic length $N_0$ which gives the scale of topological effects.  
Thus, when entropic repulsions are balanced with the knot complexity for a composite knot $K_1 \# K_2$,  the equilibrium length $N_{12}$ is smaller in RP than in SAP, and hence we have smaller fractions $N_{12}/(N_1 + N_2)$.

%%%%%%%%%%%%%%%%%%%%%%%%%%%%%%%%%%%%%%%%%%%%%%%%%%%%%%%%%%
%
% Sec 7
%
%%%%%%%%%%%%%%%%%%%%%%%%%%%%%%%%%%%%%%%%%%%
\section{Concluding remarks}

For topological swelling of the RP with a fixed knot $K$ we have argued that the topological finite-size effect plays a central role in the mean-square radius of gyration of the RP under the topological constraint $K$ plotted against segment number $N$. 

We have  shown  that the three-parameter formula (\ref{eq:RK}) with scaling exponent as a fitting parameter gives a good fitted curve to the plot of the mean-square gyration radius of the cylindrical SAP with a fixed knot against segment number $N$ for any given value of radius $r_{\rm ex}$.  
The results should be useful for describing the mean-square radius of gyration for knotted semi-flexible ring polymers.  

We have argued that the effective scaling exponent of the RP with a knot $K$ is given by 
the scaling exponent of SAW  $\nu_{\rm SAW}$ if the upper limit of the plot range of $N$ becomes infinitely large such as $N=10^8$ or larger than that.    
However, if the upper limit of $N$ is given by a finite number such as $N=10^4$, 
it is definitely less than  $\nu_{\rm SAW}$ with respect to errors. 

It follows that, at least for segment numbers $N$ less than $10^4$, to the plots of the mean-square radius of gyration for the RP under a topological constraint against segment number $N$ it is impossible to apply such a large-$N$ asymptotic expansion that would have the effective scaling exponent equal to the scaling exponent of SAW and give good theoretical curves with small $\chi^2$ values where there are only three fitting parameters.  

We have shown that the equilibrium length of a composite knot is approximately 
equal to  the sum of those of constituent prime knots of which the composite knot consists.  Furthermore, we have argued that the additivity of equilibrium lengths for composite knots is compatible with the local knot conjecture.

\section*{Acknowledgements} 
The authors would like to thank T. Prellberg for helpful comments. 
They are also thankful for comments to many participants in the Conference on Means, Methods and Results in the Statistical Mechanics of Polymeric Systems II, June 12 -14, 2017, The Fields Institute, Toronto, Canada.  
The present research is partially supported by the Grant-in-Aid for Scientific Research No. 26310206.

\newpage
\appendix

\setcounter{equation}{0} 
\setcounter{figure}{0} 
\setcounter{table}{0} 
\renewcommand{\theequation}{S.\arabic{equation}}
\renewcommand{\thefigure}{S\arabic{figure}}
\renewcommand{\thetable}{S\arabic{table}}
\renewcommand{\thesection}{S\arabic{section}}

\section{Ratio of topological swelling for the figure-eight knot} 
%$4_1$}
\label{secA}

%-----------------------------------------------------------------------
% Figure   Radius of gyration 41
%
\begin{figure}[htbp]
\begin{center}
  \includegraphics[width=
%11.0cm 
0.7\hsize
]{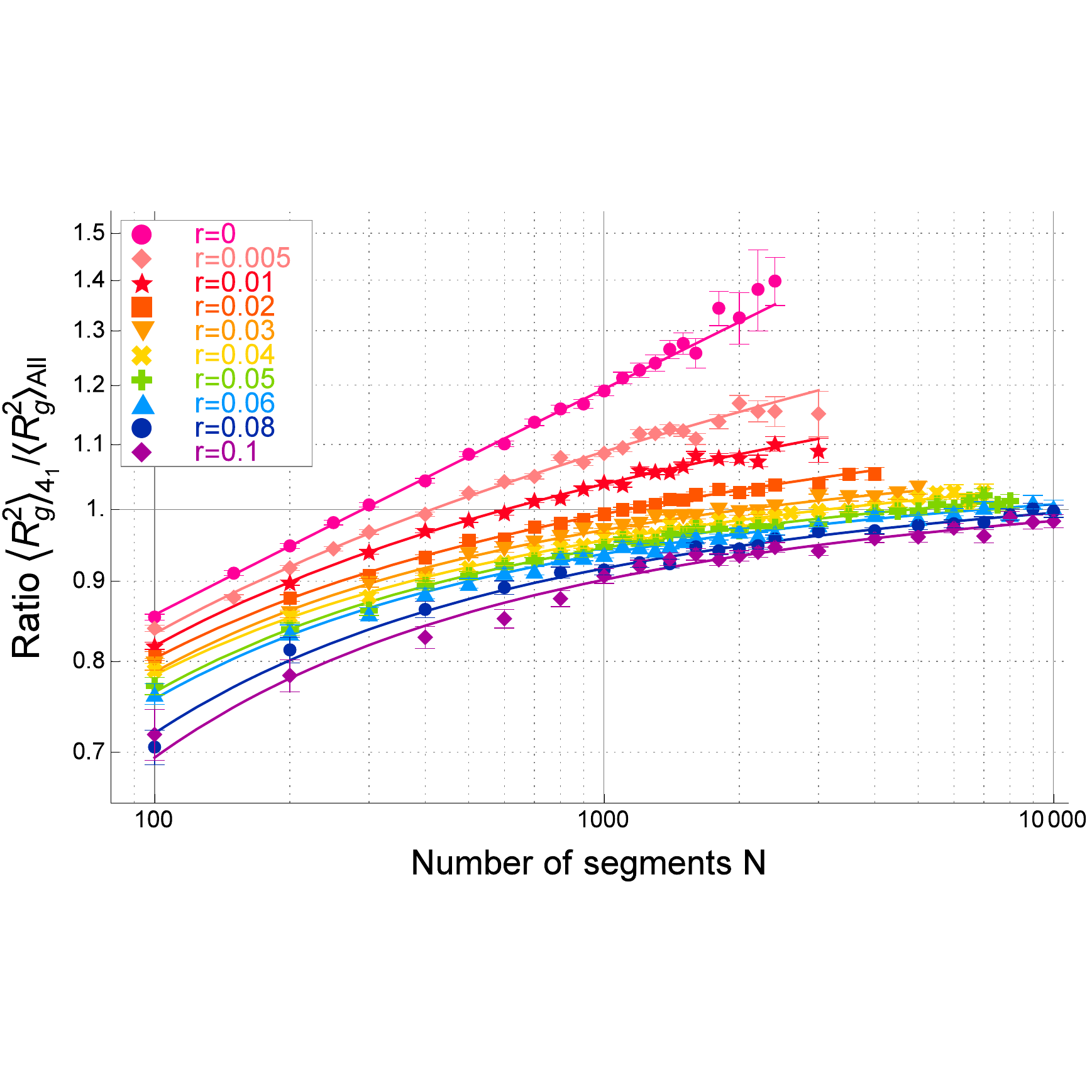}
  \caption{Ratio of  topological swelling for the figure-eight knot ($4_1$)
in the cylindrical SAP with radius  $r_{\rm ex}$,   
$\langle R^2_g \rangle_{4_1}/\langle R^2_{g} \rangle_{All}$,  
 plotted against the number of segments $N$ in the double-logarithmic scale. }
  \label{fig:ratio41}
\end{center}
\end{figure}
%-----------------------------------------------------------------------

\begin{table}[htbp]
\begin{center} 
\begin{tabular}{c|cccc}  \hline 
$r_{ex}$ & $a_K$ & $\Delta \nu_K$ & $b_K$ & $\chi^2/{\rm DF}$ \\
 \hline 
0 & $0.454 \pm 0.026$ & $0.0702 \pm 0.0036$ & $-0.12 \pm 0.25$ & $0.69$  \\ 
0.005 & $0.753 \pm 0.044$ & $0.0306 \pm 0.0037$ & $-1.65 \pm 0.24$ & $1.81$  \\ 
0.01 & $0.869 \pm 0.037$ & $0.0175 \pm 0.0027$ & $-1.99 \pm 0.18$ & $1.31$  \\ 
0.02 & $0.913 \pm 0.026$ & $0.0110 \pm 0.0017$ & $-2.04 \pm 0.14$ & $0.89$  \\ 
0.03 & $0.981 \pm 0.030$ & $0.0047 \pm 0.0018$ & $-2.31 \pm 0.16$ & $1.24$  \\ 
0.04 & $0.936 \pm 0.022$ & $0.0065 \pm 0.0013$ & $-2.11 \pm 0.13$ & $0.75$  \\ 
0.05 & $0.963 \pm 0.025$ & $0.0043 \pm 0.0014$ & $-2.37 \pm 0.16$ & $0.95$  \\ 
0.06 & $0.960 \pm 0.025$ & $0.0039 \pm 0.0015$ & $-2.40 \pm 0.17$ & $0.81$  \\ 
0.08 & $0.942 \pm 0.031$ & $0.0044 \pm 0.0018$ & $-2.66 \pm 0.22$ & $0.67$  \\ 
0.1 & $0.930 \pm 0.049$ & $0.0046 \pm 0.0029$ & $-2.84 \pm 0.37$ & $1.00 $  \\ 
 \hline 
\end{tabular}
\end{center} 
\caption{Best estimates of the parameters in eq (\ref{eq:ratioRK}) for the ratio of the mean-square radius of gyration for the cylindrical SAP with the figure-eight knot $4_1$ to that of no topological constraint, denoted by 
$\langle R^2_g \rangle_{4_1}/\langle R^2_{g} \rangle_{All}$.}
\label{tab:ratio-41}.
\end{table}

\newpage
\section{Gyration radius of SAP with a fixed knot}
\label{secB}

%%%%%%%%%%%%%%%%%%%
%\input{table-RK.tex}
%%%%%%%%%%%%%%%%%%%

\begin{table}[htbp] 
\begin{center}
\begin{tabular}{c|cccc} 
 \hline $r_{\rm ex}$ & $A_{K}$ & $\nu_K$ & $B_{K}$ & $\chi^2/{\rm DF}$ \\ \hline 
\multicolumn{5}{c}{Knot $0_1$} \\ \hline 
0 & $0.0342 \pm 0.0013$ & $0.5743 \pm 0.0024$ & $3.17 \pm 0.22$ & $0.72$  \\ 
0.005 & $0.0376 \pm 0.0010$ & $0.5749 \pm 0.0017$ & $2.41 \pm 0.15$ & $1.00$  \\ 
0.01 & $0.03779 \pm 0.00073$ & $0.5794 \pm 0.0012$ & $2.30 \pm 0.11$ & $0.83$  \\ 
0.02 & $0.03987 \pm 0.00046$ & $0.5837 \pm 0.0007$ & $1.931 \pm 0.066$ & 
$0.73$  \\ 
0.03 & $0.04154 \pm 0.00051$ & $0.58710 \pm 0.00073$ & $1.709 \pm 0.071$ & 
$1.59$  \\ 
0.04 & $0.04447 \pm 0.00033$ & $0.58763 \pm 0.00043$ & $1.419 \pm 0.043$ & 
$1.07$  \\ 
0.05 & $0.04697 \pm 0.00023$ & $0.58841 \pm 0.00028$ & $1.197 \pm 0.029$ & 
$0.70$  \\ 
0.06 & $0.05029 \pm 0.00027$ & $0.58783 \pm 0.00031$ & $0.928 \pm 0.032$ & 
$1.20$  \\ \hline 
%\end{tabular}  \caption{$0_1$} \label{tab:RK01} \end{table} 
% \begin{table}[htbp] \begin{tabular}{c|cccc} \hline 
%$r_{\rm ex}$ & $A_{K}$ & $\nu_K$ & $B_{K}$ & $\chi^2/{DF}$ \\ \hline 
\multicolumn{5}{c}{Knot $3_1$} \\ \hline 
0 & $0.0421 \pm 0.0013$ & $0.5635 \pm 0.0020$ & $0.25 \pm 0.14$ & $0.92$  \\ 
0.005 & $0.0408 \pm 0.0010$ & $0.5715 \pm 0.0016$ & $0.25 \pm 0.13$ & $1.15$  \\ 
0.01 & $0.04109 \pm 0.00077$ & $0.5761 \pm 0.0011$ & $0.106 \pm 0.097$ & 
$0.89$  \\ 
0.02 & $0.04301 \pm 0.00086$ & $0.5807 \pm 0.0012$ & $-0.21 \pm 0.11$ & 
$1.70$  \\ 
0.03 & $0.04565 \pm 0.00050$ & $0.5827 \pm 0.00064$ & $-0.490 \pm 0.065$ & 
$0.76$  \\ 
0.04 & $0.04839 \pm 0.00037$ & $0.58391 \pm 0.00044$ & $-0.779 \pm 0.049$ & 
$0.57$  \\ 
0.05 & $0.05091 \pm 0.00053$ & $0.5849 \pm 0.00057$ & $-1.026 \pm 0.070$ & 
$1.19$  \\ 
0.06 & $0.05269 \pm 0.00044$ & $0.58621 \pm 0.00046$ & $-1.154 \pm 0.060$ & 
$0.85$  \\ \hline 
%\end{tabular} \caption{$3_1$} \label{tab:RK31} \end{table} 
% \begin{table}[htbp] \begin{tabular}{c|cccc}  \hline 
%$r_{\rm ex}$ & $A_{K}$ & $\nu_K$ & $B_{K}$ & $\chi^2/{DF}$ \\ \hline 
\multicolumn{5}{c}{Knot $4_1$} \\ \hline 
0 & $0.0368 \pm 0.0022$ & $0.5717 \pm 0.0038$ & $0.13 \pm 0.27$ & $0.69$  \\ 
0.005 & $0.0394 \pm 0.0025$ & $0.5741 \pm 0.0040$ & $-0.46 \pm 0.30$ & $1.64$  \\ 
0.01 & $0.0401 \pm 0.0021$ & $0.5776 \pm 0.0033$ & $-0.63 \pm 0.26$ & $1.46$  \\ 
0.02 & $0.0404 \pm 0.0014$ & $0.5846 \pm 0.0021$ & $-0.76 \pm 0.19$ & $0.86$  \\ 
0.03 & $0.0443 \pm 0.0017$ & $0.5845 \pm 0.0022$ & $-1.22 \pm 0.21$ & $1.27$  \\ 
0.04 & $0.0438 \pm 0.0012$ & $0.5895 \pm 0.0015$ & $-1.11 \pm 0.17$ & $0.76$  \\ 
0.05 & $0.0475 \pm 0.0013$ & $0.5887 \pm 0.0015$ & $-1.59 \pm 0.18$ & $0.93$  \\ 
0.06 & $0.0501 \pm 0.0013$ & $0.5889 \pm 0.0014$ & $-1.78 \pm 0.17$ & $0.75$  \\ 
\hline 
\end{tabular} 
\end{center} 
\caption{Best estimates of the parameters of eq (\ref{eq:RK}) for 
knots $0_1$, $3_1$ and $4_1$. 
%the trivial knot $0_1$, the trefoil knot $3_1$, and the figure-eight knot $4_1$. 
}
\label{tab:RK034}
\end{table}

\begin{table}[htbp] 
\begin{center}
\begin{tabular}{c|cccc} 
 \hline 
$r_{\rm ex}$ & $A_{K}$ & $\nu_K$ & $B_{K}$ & $\chi^2/{\rm DF}$ \\ \hline
\multicolumn{5}{c}{Knot $5_1$} \\ \hline 
0 & $0.0371 \pm 0.0049$ & $0.5706 \pm 0.0084$ & $-0.62 \pm 0.56$ & $1.38$  \\ 
0.005 & $0.0373 \pm 0.0034$ & $0.5768 \pm 0.0057$ & $-0.75 \pm 0.41$ & 
$1.05$  \\ 
0.01 & $0.0315 \pm 0.0028$ & $0.5917 \pm 0.0054$ & $0.04 \pm 0.46$ & $0.93$  \\ 
0.02 & $0.0395 \pm 0.0030$ & $0.5858 \pm 0.0046$ & $-1.34 \pm 0.38$ & $1.13$  \\ 
0.03 & $0.0426 \pm 0.0032$ & $0.5871 \pm 0.0044$ & $-1.79 \pm 0.40$ & $1.13$  \\ 
0.04 & $0.0408 \pm 0.0027$ & $0.5938 \pm 0.0037$ & $-1.56 \pm 0.41$ & $0.93$  \\ 
0.05 & $0.0434 \pm 0.0030$ & $0.5936 \pm 0.0038$ & $-1.66 \pm 0.47$ & $0.86$  \\ 
0.06 & $0.0415 \pm 0.0026$ & $0.5991 \pm 0.0034$ & $-1.48 \pm 0.47$ & $0.58$  \\ 
\hline 
%\end{tabular} \caption{$5_1$}\label{tab:RK51} \end{table} 
%\begin{table}[htbp] \begin{tabular}{c|cccc}  \hline 
%$r_{\rm ex}$ & $A_{K}$ & $\nu_K$ & $B_{K}$ & $\chi^2/{DF}$ \\  \hline 
\multicolumn{5}{c}{Knot $5_2$} \\ \hline 
0 & $0.0313 \pm 0.0025$ & $0.5819 \pm 0.0052$ & $0.12 \pm 0.38$ & $0.74$  \\ 
0.005 & $0.0371 \pm 0.0022$ & $0.5769 \pm 0.0037$ & $-0.76 \pm 0.27$ & 
$0.77$  \\ 
0.01 & $0.0360 \pm 0.0013$ & $0.5843 \pm 0.0023$ & $-0.84 \pm 0.18$ & $0.35$  \\ 
0.02 & $0.0370 \pm 0.0022$ & $0.5892 \pm 0.0036$ & $-0.93 \pm 0.31$ & $1.06$  \\ 
0.03 & $0.0370 \pm 0.0018$ & $0.5949 \pm 0.0029$ & $-1.06 \pm 0.28$ & $0.72$  \\ 
0.04 & $0.0434 \pm 0.0024$ & $0.5901 \pm 0.0031$ & $-1.93 \pm 0.33$ & $1.14$  \\ 
0.05 & $0.0476 \pm 0.0020$ & $0.5882 \pm 0.0023$ & $-2.38 \pm 0.25$ & $0.70$  \\ 
0.06 & $0.0458 \pm 0.0027$ & $0.5935 \pm 0.0032$ & $-2.09 \pm 0.41$ & $1.11$  \\ 
\hline 
\end{tabular} 
\end{center}
\caption{Best estimates of the parameters of eq (\ref{eq:RK}) for  knots $5_1$ and $5_2$. }
\label{tab:RK5152}
\end{table}

\begin{table}[htbp] 
\begin{center}
\begin{tabular}{c|cccc} 
 \hline 
$r_{\rm ex}$ & $A_{K}$ & $\nu_K$ & $B_{K}$ & $\chi^2/{\rm DF}$ \\ \hline 
\multicolumn{5}{c}{Knot $3_1 \# 3_1$} \\ \hline 
0 & $0.0404 \pm 0.0020$ & $0.5667 \pm 0.0031$ & $-0.78 \pm 0.23$ & $1.24$  \\ 
0.005 & $0.0422 \pm 0.0014$ & $0.5704 \pm 0.0021$ & $-1.17 \pm 0.17$ & 
$1.09$  \\ 
0.01 & $0.0415 \pm 0.0016$ & $0.5762 \pm 0.0022$ & $-1.19 \pm 0.21$ & $1.42$  \\ 
0.02 & $0.0438 \pm 0.0015$ & $0.580 \pm 0.002$ & $-1.47 \pm 0.22$ & $1.41$  \\ 
0.03 & $0.0458 \pm 0.0013$ & $0.5831 \pm 0.0015$ & $-1.79 \pm 0.19$ & $0.99$  \\ 
0.04 & $0.0500 \pm 0.0011$ & $0.5826 \pm 0.0012$ & $-2.22 \pm 0.17$ & $0.83$  \\ 
0.05 & $0.0521 \pm 0.0015$ & $0.5841 \pm 0.0015$ & $-2.48 \pm 0.24$ & $1.19$  \\ 
0.06 & $0.0542 \pm 0.0014$ & $0.5851 \pm 0.0013$ & $-2.70 \pm 0.23$ & $0.79$  \\ 
\hline 
%\end{tabular} \caption{$3_1\# 3_1$} \label{tab:RK3131} \end{table} 
%\begin{table}[htbp] \begin{tabular}{c|cccc}  \hline 
%$r_{\rm ex}$ & $A_{K}$ & $\nu_K$ & $B_{K}$ & $\chi^2/{DF}$ \\ \hline 
\multicolumn{5}{c}{Knot $3_1 \# 4_1 $} \\ \hline 
0 & $0.0382 \pm 0.0026$ & $0.5707 \pm 0.0042$ & $-1.29 \pm 0.30$ & $0.99$  \\ 
0.005 & $0.0395 \pm 0.0013$ & $0.5744 \pm 0.0020$ & $-1.54 \pm 0.16$ & $0.40$  \\ 
0.01 & $0.038 \pm 0.002$ & $0.5814 \pm 0.0030$ & $-1.44 \pm 0.28$ & $0.87$  \\ 
0.02 & $0.0399 \pm 0.0022$ & $0.5856 \pm 0.0031$ & $-1.74 \pm 0.33$ & $0.95$  \\ 
0.03 & $0.0385 \pm 0.0020$ & $0.5930 \pm 0.0029$ & $-1.41 \pm 0.37$ & $0.81$  \\ 
0.04 & $0.0407 \pm 0.0025$ & $0.5939 \pm 0.0033$ & $-1.71 \pm 0.49$ & $1.17$  \\ 
0.05 & $0.0482 \pm 0.0026$ & $0.5885 \pm 0.0028$ & $-2.89 \pm 0.43$ & $0.86$  \\ 
0.06 & $0.0513 \pm 0.0029$ & $0.5880 \pm 0.0029$ & $-3.17 \pm 0.49$ & $0.75$  \\ 
\hline 
\end{tabular} 
\end{center}
\caption{Best estimates of the parameters of eq (\ref{eq:RK}) for the composite knots $3_1\# 3_1$and  $3_1\# 4_1$. }
\label{tab:RK3334}
\end{table}

%313131knots

\begin{table}[htbp]
\begin{center}
\begin{tabular}{c|cccc}
 \hline 
$r_{\rm ex}$ & $A_{K}$ & $\nu_K$ & $B_{K}$ & $\chi^2/{\rm DF}$ \\ \hline 
0 & $0.0467 \pm 0.0026$ & $0.5589 \pm 0.0034$ & $-2.67 \pm 0.25$ & $0.97$  \\
0.005 & $0.0393 \pm 0.0021$ & $0.5756 \pm 0.0031$ & $-2.13 \pm 0.28$ & $0.96$  \\
0.01 & $0.0442 \pm 0.0028$ & $0.5733 \pm 0.0036$ & $-2.71 \pm 0.37$ & $1.39$  \\
0.02 & $0.0464 \pm 0.0022$ & $0.5778 \pm 0.0026$ & $-3.10 \pm 0.33$ & $1.00$  \\
0.03 & $0.0454 \pm 0.0030$ & $0.5839 \pm 0.0035$ & $-2.63 \pm 0.53$ & $1.45$  \\
0.04 & $0.0522 \pm 0.0030$ & $0.581 \pm 0.003$ & $-3.87 \pm 0.48$ & $1.23$  \\
0.05 & $0.0525 \pm 0.0035$ & $0.5841 \pm 0.0034$ & $-3.71 \pm 0.62$ & $1.02$  \\
0.06 & $0.0563 \pm 0.0040$ & $0.5834 \pm 0.0035$ & $-4.11 \pm 0.69$ & $0.80$  \\
0.08 & $0.059 \pm 0.015$ & $0.586 \pm 0.013$ & $-4.0 \pm 2.8$ & $1.17$  \\ 
0.1 & $0.037 \pm 0.02$ & $0.613 \pm 0.026$ & $1.8 \pm 7.3$ & $0.65$  \\
\hline 
\end{tabular}
\end{center}
\caption{Best estimates of the parameters of eq (\ref{eq:RK}) for  knot $3_1 \# 3_1 \# 3_1$. }
\label{tab:RK313131}
\end{table}

%313141knots

\begin{table}[htbp]
\begin{center}
\begin{tabular}{c|cccc}
 \hline 
$r_{\rm ex}$ & $A_{K}$ & $\nu_K$ & $B_{K}$ & $\chi^2/{\rm DF}$ \\ \hline 
0 & $0.0357 \pm 0.0029$ & $0.5751 \pm 0.0048$ & $-2.02 \pm 0.39$ & $0.96$  \\
0.005 & $0.0370 \pm 0.0035$ & $0.5783 \pm 0.0055$ & $-2.15 \pm 0.51$ & $1.60$  \\
0.01 & $0.0431 \pm 0.0030$ & $0.5746 \pm 0.0040$ & $-3.12 \pm 0.40$ & $0.84$  \\
0.02 & $0.0437 \pm 0.0029$ & $0.5811 \pm 0.0036$ & $-3.39 \pm 0.45$ & $0.90$  \\
0.03 & $0.0463 \pm 0.0044$ & $0.583 \pm 0.005$ & $-3.66 \pm 0.72$ & $1.14$  \\
0.04 & $0.0434 \pm 0.0043$ & $0.5905 \pm 0.0051$ & $-2.91 \pm 0.87$ & $0.86$  \\
0.05 & $0.0540 \pm 0.0055$ & $0.5827 \pm 0.0052$ & $-4.77 \pm 0.90$ & $0.81$  \\
0.06 & $0.0471 \pm 0.0072$ & $0.5929 \pm 0.0076$ & $-3.8 \pm 1.5$ & $0.96$  \\
0.08 & $0.089 \pm 0.019$ & $0.567 \pm 0.011$ & $-10.45 \pm 0.94$ & $2.23$  \\
%0.1 & $0.003 \pm 0.085$ & $0.7 \pm 1.3$ & $100. \pm 1100.$ & $2.18$  \\
\hline 
\end{tabular}
\end{center}
\caption{Best estimates of the parameters of eq (\ref{eq:RK}) for  knot $3_1 \# 3_1 \# 4_1$. }
\label{tab:RK313141}
\end{table}

%314141knots

\begin{table}[htbp]
\begin{center}
\begin{tabular}{c|cccc}
 \hline 
$r_{\rm ex}$ & $A_{K}$ & $\nu_K$ & $B_{K}$ & $\chi^2/{\rm DF}$ \\ \hline 
0 & $0.0403 \pm 0.0041$ & $0.5670 \pm 0.0062$ & $-3.12 \pm 0.44$ & $0.48$  \\
0.005 & $0.0240 \pm 0.0044$ & $0.604 \pm 0.011$ & $-0.40 \pm 1.2$ & $0.85$  \\
0.01 & $0.0597 \pm 0.0088$ & $0.5554 \pm 0.0086$ & $-5.33 \pm 0.72$ & $1.13$  \\
0.02 & $0.0338 \pm 0.0061$ & $0.5955 \pm 0.0097$ & $-2.5 \pm 1.3$ & $0.67$  \\
0.03 & $0.0587 \pm 0.0098$ & $0.5712 \pm 0.0091$ & $-6.4 \pm 1.1$ & $1.00$  \\
0.04 & $0.051 \pm 0.012$ & $0.582 \pm 0.012$ & $-4.6 \pm 1.8$ & $0.91$  \\
0.05 & $0.0495 \pm 0.0083$ & $0.5869 \pm 0.0091$ & $-4.84 \pm 0.90$ & $0.99$  \\
0.06 & $0.166 \pm 0.053$ & $0.530 \pm 0.017$ & $-15.3 \pm 1.9$ & $0.79$  \\
0.08 & $0.070 \pm 0.049$ & $0.580 \pm 0.036$ & $-12.7 \pm 4.7$ & $0.50$  \\
%0.1 & $0. \pm 0.25$ & $0.8 \pm 8.5$ & $0. \pm 11000.$ & $0.92$  \\
\hline 
\end{tabular}
\end{center}
\caption{Best estimates of the parameters of eq (\ref{eq:RK}) for  knot $3_1 \# 4_1 \# 4_1$. }
\label{tab:RK314141}
\end{table}

\newpage 
\section{Best estimates to the plots of the mean-square gyration radius for cylindrical SAP of no topological constraint versus segment number} 
%$\langle R^2_{g} \rangle_{All}$ versus $N$}
\label{secC}

By applying eq (\ref{eq:RK}) in the case of $K=all$ 
we derive the fitted curves to the plots of the mean-square radius of gyration for the 
cylindrical SAP under no topological constraint $\langle R^2_{g} \rangle_{All}$ 
against the number of segments  $N$ for several different values of cylindrical radius $r_{\rm ex}$. 
The curves fit to the data points very well and the $\chi^2$ values per DF  are less than 2.0 for almost all the values of cylindrical radius $r_{\rm ex}$.  The best estimates of the parameters 
are listed in Table \ref{tab:gyr-all}.

Moreover, the absolute value of the coefficient of the finite-size correction term $B_{All}$ is less than 1.0 for the large values of cylindrical radius satisfying $r_{\rm ex} \ge 0.06$, as listed in Table \ref{tab:gyr-all}.   Therefore, for $N > 10^3$, the finite-size correction term is very small such as less than 3 percentages in the case of the large values of the cylindrical radius with $r_{\rm ex} \ge 0.06$.

\begin{table}[htbp] 
\begin{center} 
\begin{tabular}{c|cccc} 
 \hline 
$r_{\rm ex}$ & $A_{All}$ & $\nu_{All}$ & $B_{All}$ & $\chi^2/{\rm DF}$ \\
 \hline 
0 & $0.08106 \pm 0.00067$ & $0.50148 \pm 0.00049$ & $0.253 \pm 0.045$ & 
$1.72$  \\ 
0.005 & $0.04610 \pm 0.00071$ & $0.55084 \pm 0.0009$ & $2.138 \pm 0.098$ & 
$3.83$  \\ 
0.01 & $0.04104 \pm 0.00048$ & $0.56652 \pm 0.00067$ & $2.379 \pm 0.076$ & 
$2.15$  \\ 
0.02 & $0.04018 \pm 0.00043$ & $0.57886 \pm 0.0006$ & $2.160 \pm 0.071$ & 
$3.14$  \\ 
0.03 & $0.04279 \pm 0.00025$ & $0.58251 \pm 0.00034$ & $1.718 \pm 0.039$ & 
$1.39$  \\ 
0.04 & $0.04502 \pm 0.00018$ & $0.58505 \pm 0.00023$ & $1.460 \pm 0.027$ & 
$0.93$  \\ 
0.05 & $0.04776 \pm 0.00012$ & $0.58610 \pm 0.00014$ & $1.184 \pm 0.017$ & 
$0.47$  \\ 
0.06 & $0.0506 \pm 0.0002$ & $0.58650 \pm 0.00022$ & $0.956 \pm 0.025$ & 
$1.09$  \\ 
0.08 & $0.05553 \pm 0.00019$ & $0.58739 \pm 0.00019$ & $0.654 \pm 0.020$ & 
$0.70$  \\ 
0.1 & $0.0608 \pm 0.00026$ & $0.58738 \pm 0.00025$ & $0.351 \pm 0.026$ & 
$1.27$  \\ 
\hline 
\end{tabular} 
\end{center} 
\caption{Best estimates of the parameters in eq (\ref{eq:RK}) for $K=all$ 
to the plots of the mean-square radius of gyration of the cylindrical SAP under no topological constraint, i.e., that of  all knots,  $\langle R_g^2 \rangle_{All}$, against segment number $N$. }
\label{tab:gyr-all}
\end{table}

%***********************************************************************
% References
%***********************************************************************
\newpage 
%\input{scaling-ref.tex} 
%\input{scaling-refRV.tex} 
%%%%%%%%%%%%%%%%%%%%%%%%%

\end{document}